\documentclass[fleqn,usenatbib,referee]{mnras}
\usepackage[dvipdfmx]{graphicx}
\usepackage{pdfpages}
\usepackage{amsmath}	% Advanced maths commands
\usepackage{amssymb}	% Extra maths symbols

\usepackage{bm}
\usepackage{multirow}

\newcommand{\msun}{\thinspace M_\odot}

\title[Growth of MRI]{Growth of Magnetorotational Instability in Circumstellar Disks around Class 0 Protostars}
\author[Y.  Kawasaki et al.]{
Yoshihiro Kawasaki,$^{1}$\thanks{E-mail: kawasaki.yoshihiro.592@s.kyushu-u.ac.jp (YK)} 
Shunta Koga,$^{1}$ and 
Masahiro N. Machida$^{1}$
\\
$^{1}$Department of Earth and Planetary Sciences, Faculty of Sciences, Kyushu University, Fukuoka 819-0395, Japan
}

\begin{document}
\label{firstpage}
\pagerange{\pageref{firstpage}--\pageref{lastpage}}
\maketitle

\begin{abstract}
We investigate the possibility of the growth of magnetorotational instability (MRI) in disks around Class 0 protostars. 
We construct a disk model and calculate the chemical reactions of neutral and charged atoms, molecules and dust grains to derive the abundance of each species and the ionization degree of the disk.  
Then, we estimate the diffusion coefficients of non-ideal magnetohydrodynamics effects such as ohmic dissipation, ambipolar diffusion and the Hall effect.
Finally, we evaluate the linear growth rate of MRI in each area of the disk. 
We investigate the effect of changes in the strength and direction of the magnetic field in our disk model and 
we  adopt four different dust models to investigate the effect of dust size distribution  on the diffusion coefficients. 
Our results indicate that an MRI active region possibly exists with a weak magnetic field in a region far from the protostar where the Hall effect plays a role in the growth of MRI.
On the other hand, in all models the disk is stable against MRI in the region within $<20$\,au from the protostar on the equatorial plane. 
Since the size of the disks in the early stage of star formation is limited to $\lesssim 10$--$20$\,au, it is difficult to develop MRI-driven turbulence in such disks.
\end{abstract}


\section{Introduction}
Magnetorotational instability (MRI) controls the evolution of accretion disks around various objects such as black holes, neutron stars and protostars, 
because MRI is considered to greatly contribute to the viscous evolution of accretion disks \citep{1991ApJ...376..214B,1998RvMP...70....1B}. 
MRI plays a significant role in disks composed of highly ionized plasma such as  the disks around (super-massive) black holes and neutron stars. 
However, it is considered that MRI does not significantly impact disks composed of very weakly ionized plasma 
\citep{1996ApJ...457..355G,1997ApJ...480..344G,1999ApJ...518..848I}. 
Protoplanetary disks, which exist around Class II and III protostars, is  composed of very weakly ionized plasma, 
because the temperature and ionization  degree of the disk are very low 
\citep{1985prpl.conf.1100H}.
In such disks, non-ideal magnetohydrodynamic (MHD) effects of ohmic dissipation, ambipolar diffusion and the Hall effect  suppress the growth of MRI
\citep{1996ApJ...457..798J,1999ApJ...515..776S,1999MNRAS.307..849W,2001ApJ...552..235B, 2004MNRAS.348..355K,2004ApJ...608..509D}.
Recent studies have shown that  MRI does not fully develop in protoplanetary disks
\citep{2008ApJ...679L.131T,2013ApJ...769...76B, 2013ApJ...764...66S, 2013ApJ...775...73S, 2015ApJ...801...84G,2015ApJ...798...84B,2015MNRAS.454.1117S}, 
which is consistent with recent observations of disks around Class I and II protostars
\citep{2015ApJ...813...99F,2017ApJ...843..150F,2018ApJ...856..117F}.

Although the ionization degree of star-forming clouds is very low, the magnetic field plays an important role in the star formation process 
\citep{2012PTEP.2012aA307I,2018FrASS...5...39W,2020SSRv..216...43Z}.
An excess of angular momentum in a star-forming cloud is removed by magnetic effects such as magnetic braking 
\citep{1979MNRAS.187..337M,1990MNRAS.242..535N,1995ApJ...452..386B,1995ApJ...453..271B} and magnetically driven wind 
\citep{1968MNRAS.138..359M,1982MNRAS.199..883B,1985PASJ...37..515U,2000prpl.conf..759K,2000ApJ...528L..41T}. 
The formation of a circumstellar disk is closely tied to both  magnetic braking and the dissipation of  the magnetic field 
\citep{2010A&A...521L..56D,2012A&A...541A..35D,2010ApJ...724.1006M, 2015ApJ...801..117T,2015MNRAS.452..278T,2019MNRAS.489.1719W}. 
A protostar forms after the first core formation in a  collapsing star-forming cloud 
\citep{1969MNRAS.145..271L,2000ApJ...531..350M}. 
Then, during the very early accretion phase, the first core (remnant) evolves into the circumstellar disk, where the magnetic field dissipates 
and the angular momentum transfer owing to  magnetic braking is alleviated 
\citep{2011PASJ...63..555M,2014MNRAS.438.2278M,2021arXiv210100131X}. 
At the protostar formation epoch, the protostar has a mass of $10^{-3}$--$10^{-2}\msun$, while the mass of the circumstellar disk formed 
from the first core is $10^{-2}$--$10^{-1}\msun$ 
\citep{2006ApJ...645..381S,2008ApJ...674..997S}.
Thus, the circumstellar disk is more massive than or comparable to the protostar \citep{2010ApJ...718L..58I}. 
Therefore,  gravitational instability occurs in the disk during the early main accretion phase 
\citep{1998ApJ...508L..95B,2009MNRAS.400...13W,2010ApJ...724.1006M}. 
As a result, during the very early stage of star formation, the angular momentum of the disk is considered to be transported by gravitational torque, not by MRI-driven turbulence 
\citep{2017ApJ...835L..11T,2019ApJ...876..149M,2020ApJ...905..174A}.   

The early star formation phase has been investigated in core collapse simulations 
\citep{2012PTEP.2012aA307I}.  
In the simulations, starting from a prestellar cloud core, the evolution of the gravitationally collapsing cloud is calculated with non-ideal MHD equations.
Such simulations have also clarified the evolution of the circumstellar disk during the main accretion phase after protostar formation or sink creation 
\citep{2011MNRAS.417.1054S,2012MNRAS.422..347S,2012A&A...543A.128J,2016MNRAS.463.4246M,2017ApJ...839...69M,2019MNRAS.486.2587W}. 
It should be noted that a sink cell or sink particle with a radius of $\sim 1$--
$10$\,au is usually used instead of resolving the protostar and the inner disk region 
\citep{2014MNRAS.438.2278M}.
The sink method can speed up the simulation and makes it possible to investigate the long-term evolution of a circumstellar disk during the main accretion phase.
Core collapse simulations can successfully reproduce various phenomena observed in star-forming regions such as protostellar jets and outflow \citep{2014ApJ...796L..17M}. 
In addition, recent  simulations indicate that a rotationally supported (or Keplerian) disk forms in a very early star formation phase
\citep{2011PASJ...63..555M}, which has been confirmed in recent ALMA observations 
\citep{2017ApJ...834..178Y,2018ApJ...864L..25O,2020MNRAS.499.4394M}. 

In core collapse simulations, the effects of the magnetic field at large scale (magnetic braking and jet driving) have been precisely investigated, 
while the growth of MRI has not yet been discussed. 
To comprehensively investigate the star formation process, simulations need to cover very different size scales, from the molecular cloud core ($\sim10^4$\,au) to the protostar or disk ($<1$\,au).  
Since little attention has been paid to resolving MRI in disks,
the spatial resolution achieved to date may not be sufficient to resolve MRI in such simulations.
No evidence of MRI growth has been found in core collapse simulations
and thus, we cannot judge whether  the absence of MRI in the early star formation phase is due to the insufficient spatial resolution or the physical state of the disk.

The growth of MRI in the late  stage of  star formation has been intensively discussed, while that in the early stage has not been sufficiently investigated. 
\citet{2019ApJ...876..149M} showed that MRI can develop only in the inner disk region ($<0.1$\,au) where the dissipation of the magnetic field is not sufficient to suppress MRI. 
However, in core collapse simulations, it is difficult to resolve MRI with sufficient spatial resolution because both the cloud and disk scales need to be resolved to adequately calculate the formation and evolution of the disk during the main accretion phase, as described above. 
Meanwhile, it is important to discuss the growth of MRI to investigate the dust growth and subsequent planet formation in the disk 
\citep{2012A&A...538A.114P,2013ApJ...771...43O,2015ApJ...801...81Z,2019ApJ...885...36H}.

This study focuses on the growth of MRI during the main accretion phase of  star formation. We analytically investigate the linear growth of MRI 
in a disk regulated by gravitational instability. This paper is structured as follows. 
We describe the disk model, non-ideal MHD effects and chemical reactions in \S2.
The results are presented in \S3. We discuss the effect of the configuration of the magnetic field and the dust growth in \S4. 
A summary is  presented in \S5.

\section{Methods}
The purpose of this study is to investigate MRI growth during the main accretion phase. 
Recent simulations have shown that a disk forms in a high density region where the ionization degree is very low and the magnetic field dissipates
\citep{2015ApJ...801..117T}. 
Thus, the angular momentum of the disk cannot be efficiently transported by magnetic braking and the magnetically driven wind. 
During the main accretion phase, the mass accretion rate onto the disk is as high as $\sim10^{-5}\msun$\,yr. 
Without an efficient mechanism of angular momentum transfer, the disk becomes massive in a short time
\citep{2020ApJ...896..158T}. 
As a result, gravitational instability occurs and a non-axisymmetric structure such as spiral arms develops in the disk
\citep{2017ApJ...835L..11T}.
Then, angular momentum is transported by  gravitational torque, and a part of the gas in the disk falls onto the central protostar.  
As an indicator of disk gravitational instability, the Toomre $Q$ parameter is usually used
\citep{2010ApJ...724.1006M,2017ApJ...835L..11T,2021arXiv210100131X}. 
Thus, we model the disk with $Q$ as in our previous study, in which the temperature profile is  given. 
Then, we calculate the chemical reaction at each area of the disk to derive the conductivities and resistivities.
Finally, we determine whether MRI develops using a linear analysis.
 
\subsection{Disk Model}
\label{sec:diskmodel}
As described above, the disk should be  regulated by the Toomre $Q$ parameter during the main accretion phase (or Class 0 and I phases). 
\citet{2019MNRAS.484.2119K} proposed a disk model around Class 0 and I protostars.
We construct an alternative disk model for Class 0 and I protostars according to \citet{2019MNRAS.484.2119K}.
We explain our disk model in this subsection.

Assuming a Keplerian rotating disk, the Toomre $Q$ parameter is described as 
\begin{equation}
Q = \frac{c_s \Omega_{\rm K}}{\pi G \Sigma_{\rm disk}},
\label{eq:toomreq}
\end{equation}
where $\Sigma_{\rm disk}$ is the disk surface density, $c_s$ is the speed of sound and $\Omega_{\rm k}$ is the Keplerian velocity, described as 
\begin{equation}
\Omega_{\rm K} = \left( \frac{G M_*}{r^3} \right)^{1/2},
\end{equation}
where $M_*$ is  the protostellar mass. 
With equation~(\ref{eq:toomreq}), the surface density can be described as 
\begin{equation}
 \Sigma_{\rm disk}=\frac{c_s \Omega_K}{\pi G Q}.
 \label{eq_eval:Sigma}
\end{equation}
To reduce the parameter set, we fix the parameters $Q$ and $M_*$ as  $Q=2$ and $M_*=0.1\msun$ in this study. 
To determine $c_s$, we adopt the temperature profile used in \citet{2019MNRAS.484.2119K}  as 
\begin{equation}
  T_{\rm disk} = 150 \left( \frac{r}{1~\rm{au}}  \right)^{- \frac{3}{7}} \ \ \ {\rm K}, 
  \label{eq_eval:DiskTemp}
\end{equation}
in which the temperature is assumed to be isothermal in the vertical direction \citep[see also][]{1997ApJ...490..368C}.
Consequently, we can describe the disk surface density as 
\begin{equation}
\Sigma_{\rm disk}= 4.2 \left(\frac{r}{100 \rm{au}} \right)^{-\frac{12}{7}} \rm{g}\, {\rm cm}^{-2}.
\label{eq_eval:Sigmanon}
\end{equation}
Assuming hydrostatic equilibrium in the $z$-direction, the density is described as 
\begin{equation}
\rho_{\rm g}(r,z) = \rho_{\rm 0}(r)\exp \left(-\frac{z^2}{2h^2} \right),
\end{equation}
where $\rho_0(r)$ is the density on the midplane $\rho_0(r) = \rho_{\rm{g}}(r, z=0)$, described as
\begin{equation}
 \rho_{\rm 0}(r) = \frac{1}{\sqrt{2\pi}}\frac{\Sigma_{\rm disk}}{h},
\end{equation}
where the scale height $h$ is described as 
\begin{equation}
h = \frac{c_{\rm s}}{\Omega_{\rm K}}.
\end{equation}

It should be noted that,  in this study, we do not actually calculate the disk evolution and growth of MRI using  a multi-dimensional MHD simulation.
Instead, we analytically model the disk regulated by Toomre Q parameter to estimate  the linear growth of MRI.
We discuss caveats of our disk model in \S\ref{sec:caveat}.

\subsection{Non-ideal MHD Effects and Diffusion Coefficients}
\label{sec:nonidealMHD}
We constructed the disk model, as described in \S\ref{sec:diskmodel}.
Thus, we can estimate the electric conductivities and magnetic dissipation coefficients of the disk given the abundance of each of the species and  the ionization degree.  
In this subsection, we overview the non-ideal MHD equations in weakly ionized plasma before we derive the conductivities 
and dissipation coefficients, which are described  in the next subsection.

The electric field of the comoving coordinates  $\bm{E}^{\prime}$ can be described by 
\begin{equation}
    \bm{E}^{\prime} = \bm{E} + \frac{1}{c}\bm{v}\times\bm{B}
    \label{mhd_eq:trans_E}
\end{equation}
where $\bm{v}$ is the velocity of neutral gas, $c$ is the speed of light,   $\bm{E}$ is the electric field of the standard of rest, and $\bm{B}$ is the magnetic flux density of the standard of rest (hereafter  magnetic field strength and magnetic field). 
In comoving coordinates, the motion of a charged particle, represented by the subscript $j$, is determined by the balance between the Lorentz  and frictional forces between the charged and  neutral particles, described as  
\begin{equation}
    Q_{j}\left(\bm{E}^{\prime} + \frac{\bm{v}_{j}}{c}\times\bm{B}\right) - \gamma_{j}\rho m_{j} \bm{v}_{j} = 0,
    \label{mhd_eq:Lorents_vs_drag}
\end{equation}
where $\gamma_{j} = \langle \sigma v \rangle_{j}/(m_{n} + m_{j})$ and $\langle\sigma v\rangle_{j}$ is the momentum exchange rate 
when the charged particle collides with a neutral particle, $m_{j}$ is the mass of charged particle, 
$\bm{v}_{j}$ is the velocity relative to $\bm{v}$, $Q_{j}$ is the charge, 
$\rho$ is the density of neutral gas and $m_{n}$ is the mean mass of a neutral particle.
The ratio of the Lorentz force to the friction force is characterized by the Hall parameter  
\begin{equation}
    \beta_{j} = \frac{Q_{j} |\bm{B}|}{m_{j}c} \frac{1}{\gamma_{j}\rho}. 
\end{equation}
We solve equation~(\ref{mhd_eq:Lorents_vs_drag}) for the velocity of a charged particle $\bm{v}_{j}$.
Then, substituting $\bm{v}_{j}$ into the current density  $\bm{J}=\Sigma_{j}Q_{j}n_{j}\bm{v}_{j}$, 
where $n_{j}$ is the number density of charged particle, 
we can derive the generalized Ohm's law as
\begin{equation}
    \bm{J} = \sigma_{O}\bm{E}^{\prime}_{\parallel} + \sigma_{H}\hat{\bm{b}}\times\bm{E}^{\prime}_{\perp} + \sigma_{P}\bm{E}_{\perp}^{\prime},
    \label{mhd_eq:general_ohm}
\end{equation}
where the subscripts $\parallel$ and $\perp$ represent components parallel and  perpendicular to the magnetic field $\bm{B}$, and $\hat{\bm{b}}$ is the unit vector of the magnetic field. 
The Ohm $\sigma_O$, Hall $\sigma_H$ and Pedersen $\sigma_P$ conductivities can be described  by 
\begin{align}
    \displaystyle \sigma_O &= \frac{c}{B}\sum_{j}Q_j n_j \beta_j, \\
    \displaystyle \sigma_H &= \frac{c}{B}\sum_{j}\frac{Q_{j}n_{j}}{1+\beta^2_j},  \\
    \displaystyle \sigma_P &= \frac{c}{B}\sum_{j}\frac{n_{j}Q_{j}\beta_j}{1+\beta^2_j}, 
\end{align}
where $B=\vert \bm{B} \vert$ is adopted \citep{1999MNRAS.303..239W}. 

Finally, we solve equation~(\ref{mhd_eq:general_ohm}) for $\bm{E}^{\prime}$.
Then, using  equation~(\ref{mhd_eq:trans_E}) and Maxwell's equations, we can derive the induction equation as
\begin{equation}
    \frac{\partial\bm{B}}{\partial t} 
    = \bm{\nabla}\times(\bm{v}\times\bm{B})
      - \bm{\nabla}\times\left[\eta_{O}\bm{\nabla}\times\bm{B} 
      + \eta_{H}(\bm{\nabla}\times\bm{B})\times\hat{\bm{b}} 
      - \eta_{A}\left(\left(\bm{\nabla}\times\bm{B}\right)\times\hat{\bm{b}}\right)\times\hat{\bm{b}}\right],
\label{eq:induction}
\end{equation}
where the coefficients of ohmic dissipation $\eta_O$, the Hall effect $\eta_H$ and ambipolar diffusion $\eta_{A}$ are described by
\begin{align}
    \eta_O &= \frac{c^2}{4\pi\sigma_O}, \\ 
    \eta_H &= \frac{c^2\sigma_H}{4\pi\sigma^2_\perp}, \\
    \eta_A &= \frac{c^2\sigma_P}{4\pi\sigma^2_\perp} - \eta_O,
\end{align}
with  $\sigma_{\perp} \equiv \sqrt{\sigma_{H}^2 + \sigma_{P}^{2}}$. 
The diffusion coefficients ($\eta_O$, $\eta_H$, $\eta_A$) and conductivities ($\sigma_O$, $\sigma_H$, $\sigma_P$) are determined by the abundance of each charge particle in the weakly ionized plasma. 
The abundance of charged particles is determined by chemical reaction calculations, as explained in the next section. 
It should be noted that  charge neutrality 
\begin{equation}
    \sum_{j} Q_{j} n_{j} = 0,
\end{equation}
is assumed in the weakly ionized plasma (or disk). %which is used to judge whether or not  the abundance calculation is correctly executed.  

To determine the diffusion coefficients of the non-ideal MHD effects, we need to determine the momentum exchange rate coefficient $\langle \sigma v \rangle_j$. 
For collisions between ions and neutral particles $\langle\sigma v\rangle_{j, in}$, we use the following equation:
\begin{equation}
  \langle\sigma v\rangle_{j, in} = 2.0\times 10^{-9} \Bigl(\frac{m_{H}}{\mu}\Bigr)^{1/2} \quad {\rm cm}^{3}\, {\rm s}^{-1},
\label{eq:in}
\end{equation}
where $\mu = (m_{i}m_{n})/(m_{i} + m_{n})$ is the reduced mass , $m_{i}$ is the ion mass and $m_{\rm{H}}$ is the hydrogen
mass \citep{2011ApJ...739...50B}.

For collisions between electrons and neutral particles $\langle \sigma v \rangle_{j,en}$, we use   
\begin{equation}
  \langle \sigma v \rangle_{j,en} = 8.3 \times 10^{-9} \times \max \left[1, \left( \frac{T}{100~\rm{K}} \right)^{1/2}\right] \quad {\rm cm}^{3}\, {\rm s}^{-1},
\label{eq:en}
\end{equation}
which is described in \citet{1983ApJ...264..485D}.
For collisions between charged dust grains and neutral particles $\langle \sigma v \rangle_{j,dn}$, we use  
\begin{equation}
  \langle \sigma v \rangle_{j,dn} = \pi a^{2} \sqrt{\frac{8 k_B T}{\pi m_n}} \quad {\rm cm}^{3}\, {\rm s}^{-1},
\label{eq:dn}
\end{equation}
where $a$ is the dust radius \citep{2011ApJ...739...50B}.

\subsection{Chemical Reactions}
\label{sec:reactions}
As described in the previous section, we need to calculate the chemical reactions to derive the electric conductivities and magnetic diffusion coefficients. 
The chemical reaction code is constructed according to \citet{2019MNRAS.484.2119K}. 
In our chemical reaction calculation, we adopt the species $\rm H$, $\rm H_{2}$, $\rm He$, $\rm C$, $\rm O$, $\rm O_{2}$, 
$\rm CO$, $\rm HCO$, $\rm H_{2}O$, $\rm OH$, $\rm Mg$, H$	^{+}$, 
$\rm H_{2}^{+}$, $\rm H_{3}^{+}$, $\rm He^{+}$, $\rm C^{+}$, $\rm O^{+}$, $\rm O_{2}^{+}$, 
$\rm H_{2}O^{+}$, $\rm H_{3}O^{+}$, $\rm OH^{+}$, $\rm O_{2}H^{+}$, $\rm CO^{+}$, $\rm HCO^{+}$, 
$\rm CH_{2}^{+}$, $\rm Mg^{+}$, and $\rm e^{-}$. 
In addition to these atoms, molecules and electrons, we include dust grains in our code because the dust abundance greatly affects 
the abundance of charged species, which determines the conductivities and resistivities 
\citep{2007Ap&SS.311...35W,2009ApJ...701..737B,2011ApJ...739...51B,2016A&A...592A..18M}. 
The dust models and properties are described in \S\ref{sec:dustmodel}. 
The reaction equation or  the evolution of the number density of a species $i$ is described as 
\begin{equation}
\frac{d n_{i}}{dt} = \sum_{j, k} k_{jk}n_{j}n_{k} + \sum_{l}k_{l}n_{l} - n_{i}\sum_{l \neq i }k_{il}n_{l} - n_{i}k_{i},
\end{equation}
where the coefficient $k_{jk}$ is the production rate of species $i$ due to the reaction of species $j$ and $k$, 
the coefficient  $k_{l}$ is the production rate of species $i$ due to the ionization of species $l$,
the coefficient $k_{il}$ is the destruction rate of species $i$ due to the reaction of species $i$ and $l$,
and $k_{i}$ is the destruction rate of species $i$ due to the ionization.

\subsubsection{Gas phase reaction}
For the gas phase reaction, the reaction coefficients are taken from the UMIST database \citep{2013A&A...550A..36M}, in which each coefficient is described as 
\begin{equation}
  k_{jk} = \alpha \Bigl(\frac{T}{300}\Bigr)^{\beta}\exp\Bigl(- \frac{\gamma}{T}\Bigr),
\end{equation}
where the values $\alpha$, $\beta$, and $\gamma$ differ in every reaction. 
Note that we exclude the reaction between cosmic rays and  protons and that between cosmic rays and photons from the UMIST database.
Alternatively, we include the reaction listed in Table~\ref{table:ionization_rate} as described in \citet{2009ApJ...701..737B}.

\subsubsection{Ionization reaction}
\label{sec:IR}
Thermal ionization does not occur in the protoplanetary disk except for the disk inner edge because the temperature is as low as $T\lesssim 2000$\,K. 
Thus, we include two ionization sources of cosmic rays and radioactive elements. 
We adopt the ionization rate due to cosmic rays described in \citet{1981PASJ...33..617U}  as
\begin{equation}
  \zeta_{\rm CR} = 1.0\times10^{-17}\Bigl\{\exp\Bigl[ -\frac{\chi}{\chi_{\rm CR}}\Bigr]
                    + \exp\Bigl[-\frac{\Sigma_{\rm disk} - \chi}{\chi_{\rm CR}}\Bigr] \Bigr\}  \ {\rm s}^{-1},
\end{equation}
where $\chi_{\rm CR} = 96\,\rm{g}\, {\rm cm}^{-2}$ is the attenuation length and $\Sigma_{\rm disk}$ is the surface density integrated from infinity $\chi = \int^{\infty}_{z} \rho_{\rm g}(r,z')dz'$. 

As the ionization due to radioactive elements, we use the ionization rate of the short-lived element $^{26}$Al \citep{2009ApJ...703.2152T} as 
\begin{equation}
  \zeta_{\rm RA} = 3.7\times 10^{-19} \  {\rm s}^{-1}.
\end{equation}
The ionization rate due to both cosmic rays and radioactive elements ($k_{l}$ and $k_{i}$) can be written as 
\begin{equation}
    k_{l}, \ k_{i} = \alpha_{\rm{ir}} \zeta,
\end{equation}
where $\zeta = \zeta_{CR} + \zeta_{RA}$ and $\alpha_{\rm{ir}}$  is the value for each reaction listed in Table~\ref{table:ionization_rate}. 

\begin{table}
  \caption{Ionization reactions. The total ionization rate $\zeta$ is described as $\zeta=\zeta_{\rm CR} + \zeta_{\rm RA}$ (for details, see \S\ref{sec:IR}).}
  \begin{tabular}{cc}
    \hline
    Reaction & Ionization Rate ($\alpha_{\rm{ir}}\zeta$) \\
    \hline
    $\rm H_2 \rightarrow H_{2}^{+} + e^{-}$ & $0.97\,  \zeta$ \\ 
    $\rm H_2 \rightarrow H^{+} + H + e^{-}$ & $0.03\,  \zeta$ \\ 
    $\rm H \rightarrow H^{+} + e^{-}$      & $0.50\, \zeta$ \\ 
    $\rm He \rightarrow He^{+} + e^{-}$    & $0.84\, \zeta$ \\
    \hline
  \end{tabular}
  \label{table:ionization_rate}
\end{table}

\subsubsection{Gas-grain reaction}
Dust grains can acquire positive and negative charges when colliding with electrons and ions. 
Dust grains that absorb electrons become negatively charged as 
\begin{equation}
    \rm g^{0} + e^{-} \rightarrow g^{-}. \notag 
\end{equation}
We assume that  the recombination of ions promptly occurs on the surface of dust grains and neutral particles escape from the grain surface 
when ions collide with dust grains \citep{2009ApJ...693.1895K}.
For example, the reaction of  dust grains and particles $\rm H$ and $\rm H^{+}$ are described as 
\begin{align}
    & \rm H^{+} + g^{-} \rightarrow H + g^{0} \notag, \\
    & \rm H^{+} + g^{+} \rightarrow H + g^{2+}  \notag. 
\end{align}
When the counterpart of ions or the corresponding neutral species are not present in the collision between ions and dust grains, dissociation reactions such as    
\begin{align}
    & \rm H_{3}^{+} + g^{-} \rightarrow H_{2} + H + g^{0} \notag, \\  
    & \rm H_{3}^{+} + g^{0} \rightarrow H_{2} + H + g^{+} \notag. 
\end{align}
are assumed to occur on the surface of dust grains, as in the dissociation reaction in the gas phase \citep{2006A&A...445..205I}: 
\begin{align}
    & \rm H_{3}^{+} + e^{-} \rightarrow H_{2} + H \notag. 
\end{align}
We adopt the collision rate between dust grains and charged particles shown in \citet{1987ApJ...320..803D}, in which 
the collision rate between electrons and charged dust grains is  
\begin{equation}
    k_{ed} = \begin{cases}
             \sigma_{\rm{d}} \sqrt{\frac{8 k_B T}{\pi m_e}} 
             \exp \left(-\frac{\it{Z}}{(1+|\it{Z}|^{-1/2})\tau_d} \right) \left\{1+ \left(\frac{1}{4\tau_d+3\it{Z}} \right)^{1/2} \right\}^2 Se   \qquad  {\rm for}   \  Z > 0, \\
             
             \sigma_{\rm{d}} \sqrt{\frac{8 k_B T}{\pi m_e}} \left(1-\frac{Z}{\tau_d} \right) \left\{1+ \left(\frac{2}{\tau_s-2Z} \right)^{1/2} \right\} Se \quad   {\rm for} \ Z < 0, 
             \end{cases}
\end{equation}
the collision rate between ions and charged dust grains is
\begin{equation}
    k_{id} = \begin{cases}
             \sigma_{\rm{d}} \sqrt{\frac{8 k_B T}{\pi m_i}} \exp \left(-\frac{\it{Z}}{(1+|\it{Z}|^{-1/2})\tau_d} \right) 
               \left\{1+ \left(\frac{1}{4\tau_d+3\it{Z}} \right)^{1/2} \right\}^2 Si 
             \quad {\rm for} \ Z > 0, \\ 
             
             \sigma_{\rm{d}} \sqrt{\frac{8 k_B T}{\pi m_i}} \ \left(1-\frac{Z}{\tau_d} \right) \ \left\{1+ \left(\frac{2}{\tau_s-2Z} \right)^{1/2} \right\} Si \quad {\rm for} \ Z < 0,
             \end{cases}
\end{equation}
and the collision rate between neutral grains and charged particles (electrons and ions) is 
\begin{equation}
  k_{jd} = \sigma_{\rm{d}} \sqrt{\frac{8 k_B T}{\pi m_i}} \left\{ 1 + \left( \frac{\pi}{2\tau_{d}} \right)^{1/2} \right\},
\end{equation}
where $Z$ is the charge number, $T$ is the gas temperature, $\sigma_{\rm{d}}$ is the cross section of dust grains, 
$a$ is the radius of dust grains, $e$ is the unit of charge and  $\tau_{d}$ is the reduced temperature (corresponding to the thermal or internal energy) described as
\begin{equation}
    \tau_{d}=\frac{a k_B T}{e^2}.
\end{equation}
The adsorption probabilities of electrons $Se$ and ions $Si$ on dust grains of $Se = 0.6$ 
and $Si = 1.0$ are adopted according to \citet{2000ApJ...543..486S} and \citet{2016A&A...592A..18M}.

\subsubsection{Grain-grain Collisions}
Without considering coagulation and destruction between dust grains, 
it is considered that dust grains are neutralized when positively charged grains collide with negatively charged grains \citep{2009ApJ...693.1895K} as 
\begin{align}
    & \rm g^{+} + g^{-} \rightarrow g^{0} + g^{0} \notag. 
\end{align}
When the charge number of dust grains differs, a singly charged dust grain remains after the neutralization of the counterpart of the dust grain as 
\begin{align}
    & \rm g^{2+} + g^{-} \rightarrow g^{+} + g^{0} \notag. 
\end{align}
We adopt the coefficient of grain-grain collision described  in \citet{1990MNRAS.243..103U} and \citet{2016A&A...592A..18M} as
\begin{equation}
    k_{gg} = \pi \left( a + a^{\prime} \right )^{2} \sqrt{\frac{8 k_{B} T}{\pi m_{\rm red,g}}}
             \left( 1 - \frac{Z Z^{\prime}e^{2}}{\left(a+a^{\prime} \right) k_{B} T } \right)
\end{equation}
where  $a$ and $a^{\prime}$ are the radius of dust grains, $Z$ and $Z^{\prime}$ are the charge number of grains, 
and $m_{\rm red, g} (\equiv m_{\rm g}m^{\prime}_{\rm{g}} / (m_{\rm g} + m^{\prime}_{\rm g})$) is  the reduced mass of two dust grains. 

\subsubsection{Formation of Molecular Hydrogen on Grain Surface} 
In star forming regions, the most abundant species is molecular hydrogen $\rm H_{2}$, which is considered to form on the surface of dust grains \citep{1963ApJ...138..393G}.
If we ignore the formation of molecular hydrogen, hydrogen atoms become abundant, which is not the case in disks around protostars. 
Thus, we should consider H$_{2}$ formation on the surface of dust grains, in which we assume physical adsorption between hydrogen atoms and dust grains.

On the grain surface, we consider  the reaction $\rm H + H \rightarrow H_{2}$, with a reaction rate determined according to \citet{2002ApJ...575L..29C} of  
\begin{equation}
    k_{\rm HH,d} \ n_{\rm H}n_{\rm d} = \frac{1}{2}v_{\rm H}\sigma_{\rm d}\eta S_{\rm H} n_{\rm H}n_{\rm d},
\label{eq:khhd}
\end{equation}
where $n_{\rm H}$ is the number density of molecular hydrogen, $n_{\rm d}$ is the number density of dust grains, 
$v_{\rm H} = \sqrt{8 k_{B}T/\pi m_{\rm H}}$ is the thermal velocity of hydrogen atoms, $m_{\rm H}$ is the hydrogen mass,
$\sigma_{\rm d}$ is the cross section of dust grains, $\eta$ is the reaction probability and $S_{\rm H}$ is the absorption coefficient, described as   
\begin{align}
    S_{\rm H} &= \frac{\int_{0}^{\infty}dE E P(E) \exp(-\frac{E}{k_{B}T})}{\int_{0}^{\infty}dE E \exp(-\frac{E}{k_{B}T})}  \notag \\
         &= 2b^{2} \left[1 - b\sqrt{\pi}\exp(b^{2})\, \mathrm{erfc}(b)\right],
\label{eq:st} 
\end{align}
where $\mathrm{erfc}(x)$ is the error function \citep{2006A&A...445..205I,2009ApJ...701..737B}. 
In equation~(\ref{eq:st}),  the $P(E)$ is  described as 
\begin{equation}
P(E) = \exp \left( -\frac{E}{2\sqrt{D\Delta E_{s}}} \right),
\end{equation}
where $D$ is the dissociation energy, $\Delta E_{s}$ is the amount of energy transferred to a grain particle due to lattice vibration and $b$ is described as
\begin{equation}
b=\frac{\sqrt{D\Delta E_{s}}}{\sqrt{2}k_{B}T}. 
\end{equation}
$D$ is approximated by the binding energy of hydrogen atoms, $D = 600.0 (K)$ \citep{2013A&A...550A..36M}, 
and $\Delta E_{s} = 2.0\times 10^{-3}$ is adopted \citep{1970JChPh..53...79H}.
In equation~(\ref{eq:khhd}), the reaction probability $\eta = 0.5$ is adopted  \citep{1979ApJS...41..555H}.

\subsection{Dust Model}
\label{sec:dustmodel}
The size of dust grains greatly affects the determination of the abundance of charged particles in the calculation of the chemical reaction. 
Thus, dust grains play a crucial role in determining the electric conductivities and the diffusion coefficients of the magnetic field
\citep{2018MNRAS.478.2723Z,2019MNRAS.484.2119K}.
In this study, we consider spherical dust grains.
The internal density of dust grains of $\rho_{\rm di} = 3.0$\,g\,cm$^{-3}$ is adopted. 
The dust-to-gas mass ratio $f_{\rm dg} (= \rho_{\rm d}/\rho_{\rm g})$ is fixed as $f_{\rm dg}=0.01$, 
where $\rho_{\rm{g}}$ is the gas density and $\rho_{\rm d}$ is the dust density.
For the size distribution of dust grains, we adopt four dust models: 
(1) single-size grains of 0.035$\mu$m (hereafter, dust model d0035), (2) single-size grains of 0.1$\mu$m (hereafter, dust model d01), 
(3) MRN grain distribution (hereafter, dust model MRN) and (4) tr-MRN distribution (hereafter, dust model tr-MRN, for details see below). 

First, we detail the single-sized dust models d01 and d0035.  
For these models, we adopt two different radii of dust grains with $a = 0.035$ and 0.1\,$\mu$m as in \citet{2020ApJ...896..158T}. 
The cross section $\sigma_{\rm{d}}$, number density $n_{\rm d}$ and density $\rho_{\rm d}$ of dust grains for the single dust models are 
\begin{eqnarray}
\sigma_{\mathrm{d}} &=& \pi a^{2}, \\
n_{\rm d} &=& \frac{3 \rho_{\rm g}}{4 \pi a^3 \rho_{\rm di}} f_{\rm dg}, \label{eqn:initialdustabunsingle} \\
\rho_{\rm d} &=& f_{\rm dg} \rho_{\rm g}.
\end{eqnarray}

Next, we describe the dust model MRN.
The MRN distribution is described as 
\begin{equation}
  \frac{d n_{\rm d}}{da} = C a^{-\lambda} \ \ \ (a_{\rm min} \le a \le a_{\rm max}), 
\end{equation}
in which $\lambda=3.5$ is adopted  \citep{1977ApJ...217..425M}. 
$C$ is the coefficient and detailed below.
For this model, we define the moment for dust size $a$ as 
\begin{equation}
  \langle a^{n} \rangle = \int_{a_{\rm min}}^{a_{\rm max}} a^{n} \frac{d n_{\rm d}}{da} da 
  = \frac{C}{n + 1 - \lambda}\left(a_{\rm max}^{n+1-\lambda} - a_{\rm min}^{n + 1 - \lambda}\right).
\end{equation}
In addition, the average moment is 
\begin{equation}
  \overline{\langle a^{n} \rangle}
  = \frac{ \int_{a_{\rm min}}^{a_{\rm max}} a^{n} \frac{dn_{\rm d}}{da}da }{ \int_{a_{\rm min}}^{a_{\rm max}}\frac{dn_{\rm d}}{da} da}
  = \frac{1 - \lambda}{n + 1 - \lambda} \frac{a_{\rm max}^{n+1-\lambda} - a_{\rm min}^{n+1-\lambda}}{a_{\rm max}^{1-\lambda} - a_{\rm min}^{1-\lambda}}.
\end{equation}
For this model, we divide the dust grains in size from $a_{\rm min}$ to $a_{\rm max}$ into $N$ logarithmically
spaced bins ($N=10$ is adopted). 
The minimum $a_{s, \rm min}$ and maximum $a_{s, \rm max}$ sizes of the s-th bin are written as  
\begin{align}
   & a_{s,\rm min} = a_{\rm min}\left(\frac{a_{\rm max}}{a_{\rm min}}\right)^{\frac{s-1}{N}}, \\
   & a_{s,\rm max} = a_{\rm min}\left(\frac{a_{\rm max}}{a_{\rm min}}\right)^{\frac{s}{N}}.
\end{align}
The typical size of dust grains in the s-th bin can be described as  
\begin{equation}
  a_{s} = \overline{\langle a \rangle}_{\rm s}
  = \frac{1 - \lambda}{3 -\lambda} \frac{a_{s, \rm max}^{2-\lambda} - a_{s, \rm min}^{2 - \lambda}}{a_{s,\rm max}^{1-\lambda} - a_{s, \rm min}^{1-\lambda}}.
\end{equation}
The typical cross section of dust grains in the s-th bin is 
\begin{equation}
  \sigma_{\mathrm{d}, s} = \pi \overline{\langle a^{2} \rangle}_{s}
  = \pi \frac{1 - \lambda}{3 -\lambda} \frac{a_{s, \rm max}^{3-\lambda} - a_{s, \rm min}^{3 - \lambda}}{a_{s,\rm max}^{1-\lambda} - a_{s, \rm min}^{1-\lambda}}.
\end{equation}
The typical volume of dust grains in the s-th bin is written as  
\begin{equation}
  v_{\mathrm{d}, s} = \frac{4}{3}\pi \overline{\langle a^{3} \rangle}_{s}
  = \pi \frac{1 - \lambda}{4 -\lambda} \frac{a_{s, \rm max}^{4-\lambda} - a_{s, \rm min}^{4 - \lambda}}{a_{s,\rm max}^{1-\lambda} - a_{s, \rm min}^{1-\lambda}}.
\end{equation}
In the equations, $\overline{\langle \  \ \rangle}_{s}$ represents the average moment in the s-th bin.
The dust number density in the s-th bin is 
\begin{equation}
  n_{\mathrm{d}, s} = \int_{a_{s, \rm min}}^{a_{s, \rm max}} \frac{dn_{d}}{da}da 
  = \frac{C}{1-\lambda} \left(a_{s, \rm max}^{1-\lambda} - a_{s, \rm min}^{1-\lambda}\right),
\end{equation}
in which the dust size distribution coefficient $C$  is derived using the mass ratio of dust grains to gas  
\begin{equation}
  \rho_{\rm d} = f_{\rm dg} \rho_{\rm g},
\end{equation}
where $\rho_{\rm d}$ is derived as 
\begin{align}
  \rho_{\rm d} &= \sum_{s = 1}^{N} m_{s} n_{\mathrm{d}, s} = \sum_{s=1}^{N} \rho_{\rm{di}} v_{\mathrm{d},s} n_{\mathrm{d},s} \notag \\ 
  &= \sum_{s = 1}^{N} \frac{4\pi \rho_{\rm{di}} C}{3(4-\lambda)} a_{\rm min}^{4-\lambda}
      \left[\left(\frac{a_{\rm max}}{a_{\rm min}}\right)^{\frac{s}{N}(4-\lambda)} - 
        \left(\frac{a_{\rm max}}{a_{\rm min}}\right)^{\frac{s - 1}{N}(4-\lambda)}\right] \notag \\ 
  &= \frac{4\pi \rho_{\rm{di}} C}{3(4-\lambda)} \left(a_{\rm max}^{4 - \lambda} - a_{\rm min}^{4-\lambda}\right), 
\end{align}
and the coefficient $C$ can be described as 
\begin{equation}
  C = \frac{3(4-\lambda) f_{\rm dg} \rho_{\rm g}}{4\pi\rho_{\rm{di}}\left(a_{\rm max}^{4-\lambda} - a_{\rm min}^{4-\lambda}\right)}.
\end{equation}
For the MRN model, the minimum  $a_{\rm min}$ and maximum $a_{\rm max}$  grain size are set to  $a_{\rm min} = 0.005\, \mu $m and $a_{\rm max} = 0.25\, \mu$m, respectively. 

Finally, we explain the dust model tr-MRN.
The tr-MRN model was used in \citet{2018MNRAS.473.4868Z}, in which dust grains with a size of $\le 0.1\, \mu$m are removed from the MRN distribution while maintaining a dust-to-gas ratio of $f_{\rm dg}=0.01$. 
Thus,  in this dust model, the minimum  and maximum  grain size are set to  $a_{\rm min} = 0.1\, \mu$m and $a_{\rm max} = 0.25\, \mu$m. 
In tr-MRN, we use the same parameter set of $\lambda =3.5$ and $N = 10$ as in the MRN model. 

The maximum charge number of dust grains is set to $\pm 2$ for d0035 and MRN, 
while $\pm 4$ is adopted for d01 model and tr-MRN
\footnote{
To save  computational time, we adopt the maximum change number of $Z_{\rm max} \pm2$ ($ \vert Z_{\rm max}\vert=2$)  for small-sized dust grain models (d0035 and MRN). 
We confirmed that the maximum charge number does not significantly affect the results when it is larger than $\vert Z_{\rm max}\vert \ge 2 $.
}. 
The dust models adopted in this study are summarized in Table~\ref{tab:dust_model}. 

\begin{table}
  \caption{ Dust grain model parameters.}       
  \begin{tabular}{cccc}
  \hline
  Model & $a \ (\rm \mu m)$       & maximum charge number $Z_{\rm max}$  \\ \hline
  d0035 & $0.035$                 & $\pm 2$             \\ 
  d01   & $0.1$                   & $\pm 4$             \\
  MRN   & $0.005 \le a \le 0.25$  & $\pm 2$             \\ 
  trMRN & $0.1 \le a \le 0.25$    & $\pm 4$             \\ 
  \hline
  \end{tabular}
  \label{tab:dust_model}
\end{table}

\subsection{Initial Abundances}
We need to determine the initial abundance of each species when calculating the chemical reaction network and estimating both the conductivities and resistivities. 
As the initial abundance of chemical species, we adopt that described in \citet{2000ApJ...543..486S} and \citet{2006A&A...445..205I}. 
We describe the abundance of species $i$ to the hydrogen nuclei  as $x_{i} = n_{i} / n_{\rm H}$. 
The abundance of magnesium $\rm Mg$ was treated as a parameter in the previous studies, while we fix it to $x_{\rm Mg} = 1.0 \times10^{-8} $ for simplicity. 
The mean molecular weight of the gas is set to be constant at $\mu = 2.34$, because the abundance change due to the reaction between molecular hydrogen and helium is not significant. 

The total number density of gas particles is approximately calculated as $n_{g} = \rho_{\rm g} / (\mu m_{\rm p})$ where $\rho_{\rm g}$ 
is the gas density and $m_{\rm p}$ is the proton mass. 
The number density of hydrogen nuclei $n_{\rm H}$ is approximated by $n_{\rm H} = n_{g} / \Sigma_{i}c_{i}$ \citep{2006A&A...445..205I}, 
where $c_{i}$ is the abundance of species $i$ in the gas phase. 
Initially,  all dust grains are set to have no charge.   

Starting from the initial abundance as described above, we solve the chemical reactions and determine the abundance of the species assuming the equilibrium state.  
Then, we derive the diffusion coefficients of the magnetic field using the abundance of the charged species and grains, as described in \S\ref{sec:nonidealMHD}.

\subsection{Distribution of Magnetic Field}
\label{sec:magneticfield}
To calculate the resistivities, we need information on the magnetic field strength in the disk. 
Recent numerical \citep{2015ApJ...801..117T} and analytical \citep{2016ApJ...830L...8H} studies indicate that the disk has a nearly uniform magnetic field of $0.1$\,G in the early main accretion phase (see also \citealt{2011MNRAS.413.2767M,2019ApJ...876..149M}).
 
For simplicity, we assume a uniform magnetic field $B_0$ in the disk. 
We adopt $B_0=0.1$\,G for the fiducial model. 
We also adopt $B_0=0.01$\,G to investigate the effect of the magnetic field on the growth of MRI. 

\subsection{Criterion of MRI}
\label{sec:crit_mri}
We apply a linear analysis of MRI in local Cartesian coordinates, as done by \citet{2012MNRAS.422.2737W} and \citet{2012MNRAS.423..222P}, to determine whether MRI grows  in each area of the disk. 
We assume  an axisymmetrical thin disk penetrated by a uniform magnetic field $\bm{B} = sB\bm{e}_{z} \ (s = \pm 1)$, in which only the $z$-component of the magnetic field is adopted. 
Note that the direction of the magnetic field is controlled by the parameter $s$.
The magnetic field  is parallel to the angular momentum vector of the disk $\bm{J}_{\rm disk}$ with $s=1$, while it is anti-parallel to $\bm{J}_{\rm disk}$ with $s=-1$.
Although we assume only a Keplerian disk, the angular velocity  $\Omega\propto r^{-q}$ is  used here. 
When  an axisymmetrical perturbation of $\propto \exp{(\sigma t - ikz)}$ is added to the disk, the dispersion relation can be described as  
\begin{equation}
  \sigma^{4} + C_{3}\sigma^{3} + C_{2}\sigma^{2} + C_{1}\sigma + C_{0} = 0,
\label{eq:dispersion}
\end{equation}
in which the coefficients $C_{3}$, $C_{2}$, $C_{1}$ and $C_{0}$ are described as
\begin{eqnarray}
   C_{3} &=& 2\left(\eta_{O} + \eta_{A}\right)k^{2}, \\ 
  C_{2} &=& \left\{ \left(\eta_{O} + \eta_{A}\right)^{2} + \eta_{H}^{2} \right\}k^{4}
    + \left(2v_{A}^{2} - q\Omega s\eta_{H}\right)k^{2} + \kappa^{2}, \\ 
  C_{1} &=& 2\left(\eta_{O} + \eta_{A}\right)v_{A}^{2}k^{4} + 2\left(\eta_{O} + \eta_{A}\right)\kappa^{2} k^{2}, \\ 
  C_{0} &=& \left[\kappa^{2}\left\{ \left( \eta_{O} + \eta_{A}\right)^{2} + \eta_{H}^{2} \right\} 
    + \left(4 - q\right)\Omega s\eta_{H} v_{A}^{2} + v_{A}^{4}\right] k^{4} - \left(2q\Omega^{2}v_{A}^2 + q\Omega s\eta_{H} \kappa^{2} \right) k^{2} \label{eq:c0},
\end{eqnarray}
where $v_{A} = B / \sqrt{4\pi\rho}$ is the Alfv\'en velocity and $\kappa^{2} = 4\Omega^{2} + d\Omega^{2}/d\ln{r} $ is the epicyclic frequency. 
The linear instability can be discussed around $\sigma = 0$ \citep{2012MNRAS.422.2737W,2012MNRAS.423..222P}, 
and the instability condition can be obtained with  $C_{0} < 0$, which corresponds to 
\begin{equation}
 \left[\kappa^{2}\left\{ \left( \eta_{O} + \eta_{A}\right)^{2} + \eta_{H}^{2} \right\} 
    + \left(4 - q\right)\Omega s\eta_{H} v_{A}^{2} + v_{A}^{4}\right] k^{4}  - \left(2q\Omega^{2}v_{A}^2 + q\Omega s\eta_{H} \kappa^{2} \right) k^{2} < 0.
\label{eq:condition}
\end{equation}
We adopt the instability condition equation~(\ref{eq:condition}) for our disk model.
Since the disk is finite, the scale of MRI is limited by the finite disk thickness or disk scale height $h$. 
Thus, the unstable wavelength is limited to $kh > 1$. 
As a result, an unstable mode of MRI exists in a local area of the disk when the conditions $\sigma > 0$ and $kh >0$ are fulfilled, and can be described by  
\begin{equation}
    \left(\frac{s\eta_{H}\Omega}{v_{A}^2} + \frac{5}{4} - \frac{3}{8}\beta\right)^{2} 
    + \left(\frac{\eta_{O}\Omega}{v_{A}^2} + \frac{\eta_{A}\Omega}{v_{A}^2}\right)^2
    < \frac{9}{16}\left(1 + \frac{1}{2}\beta\right)^{2}, 
    \label{eq_eval:criterion_mri}
\end{equation}
where $\beta = 2 c_{s}^{2} / v_{A}^2$ is the plasma beta.
Equation~(\ref{eq_eval:criterion_mri}) is equivalent to equation~(27) of \citet{2012MNRAS.422.2737W}. 
We can estimate  the diffusion coefficients $\eta_{O}$, $\eta_{A}$ and $\eta_{H}$ from our disk model. 
Using these  coefficients, we evaluate whether the unstable condition of MRI is fulfilled with the dispersion relation equation~(\ref{eq_eval:criterion_mri}). 
When the unstable condition is fulfilled, the growth rate of the fastest growing mode of MRI is driven (for details, see also Appendix B of \citealt{2012MNRAS.422.2737W}). 
It should be noted that above conditions are valid when $kh \gg 1$, while we extend to mode of $kh\sim 1$ to derive the most unstable wavelength.
Thus, the growth rates of MRI shown in \S\ref{sec:results} are somewhat different from the actual rates. 

In the following, we simply describe the instabilities and waves derived from the dispersion relation.
Firstly, taking  the  $\eta_O$,  $\eta_H$, $\eta_A \rightarrow 0$ limit, equations~(\ref{eq:dispersion})--(\ref{eq:c0}) correspond to the dispersion relation for the standard MRI without diffusion terms ($\eta_0$  $\eta_H$, $\eta_A$).
In this case  (the standard MRI case), the dispersion relation can be written as  
\begin{equation}
  \sigma^{4} + \left(2 v_{A}^{2} k^{2} + \kappa^{2}\right)\sigma^{2} 
    + k^{2} v_{A}^{2}\left( k^{2}v_{A}^{2} - 2q\Omega^{2} \right) = 0,
\end{equation}
and  the instability condition is  described as $ k^{2}v_{A}^{2} - 2q\Omega^{2} < 0$.

Secondly, we consider the instability condition when non-ideal MHD terms are included. 
Both ohmic dissipation and ambipolar diffusion tend to stabilize MRI, while the Hall effect can promote the growth of other two instabilities (Hall-shear instability and diffusive MRI). 
Before describing  these instabilities, we introduce two different waves caused by the Hall effect.
Adopting $\eta_{O}=0$ and $\eta_{A} = 0$ and taking the $\Omega\rightarrow 0$ limit (i.e. no rotation), the dispersion relation 
(eqs.~[\ref{eq:dispersion}]--[\ref{eq:c0}]) can be written as 
\begin{equation}
  \sigma^{4} + \left( \eta_{H}^{2}k^{4} + 2k^{2}v_{A}^{2} \right)\sigma^{2} + k^{4}v_{A}^{4} = 0.
\label{eq:sigma2}
\end{equation}
For equation~(\ref{eq:sigma2}),  we can obtain the following two solutions with $\sigma = i\omega$,  
\begin{equation}
  \omega = \pm \frac{\eta_{H} k^{2}}{2} + \sqrt{\frac{\eta_{H}^{2}k^{4}}{4} + k^{2}v_{A}^{2}}.
\label{eq:omega}
\end{equation}
Taking the  $k \rightarrow 0$ limit  (i.e. short wavelength limit), equation~(\ref{eq:omega}) corresponds to the Alfv\'en wave. 
On the other hand, when the wavenumber $k$ is large, there are two branches (corresponding to whistler and ion-cyclotron waves) 
depending on the sign of the first term ($\pm \eta_{H} k^{2}/2$) in the right-hand side. 
With the positive sign ($+\eta_{H} k^{2}$),  whistler wave appears as $\omega\simeq\eta_{H}k^{2}$. 
With the minus sign ($-\eta_{H} k^{2}$),  we have ion-cyclotron wave $\omega \simeq  v_{A}^{2}/\eta_{H}$ as $k$ increases.

Finally, we consider the case of  differential rotation.
For the dispersion relation of equations~(\ref{eq:dispersion})--(\ref{eq:c0}), we take  the limit of $k v_{A} \rightarrow 0$ 
while maintaining the condition $|\eta_{H}k| > 0$.
In this case, the dispersion relation can be described  as 
\begin{equation}
  \left(\sigma^{2} + \kappa^{2}\right)
  \left( \sigma^{2} + \eta_{H}^{2}k^{4} - s \eta_{H} q\Omega k^{2} \right) = 0.
\label{eq:dispersion3}
\end{equation}
With this procedure, the induction  equation is decoupled from the equation of motion, and the evolution of magnetic field (or magnetic flux density) 
is determined only by the diffusion (term) \citep{2005A&A...434..629R}. 
Using equation~(\ref{eq:dispersion3}), the instability condition is  written  as 
\begin{equation}
\eta_{H}^{2}k^{2} - s\eta_{H}q\Omega < 0.
\label{eq:condition1}
\end{equation}
Thus, to realize  the instability condition of equation~(\ref{eq:condition1}), the necessary condition (not necessary and sufficient condition) is $s\eta_{H} > 0$. 
This instability is called as Hall-shear instability \citep{2008MNRAS.385.1494K} which 
can grow when the magnetic perturbation of whistler wave develops due to rotational shear.

Then, we consider the $k\rightarrow \infty$ limit. 
In this case, the frequency of whistler wave approaches infinity, while that of ion-cyclotron wave is finite. 
Assuming the growth rate of the perturbation is finite, the dispersion relation can be described as 
\begin{equation}
  \eta_{H}^{2}k^{4}\sigma^{2} + \left[ k^{2}v_{A}^{2} + \left(2 - \Omega\right)s\eta_{H}\Omega k^{2} \right]
    \left[ k^{2}v_{A}^{2} + 2s\eta_{H}\Omega k^{2} \right] = 0,
\label{eq:dispersion4}
\end{equation}
in which only terms with $\mathcal{O}(k^4)$ remain in equations~(\ref{eq:dispersion})--(\ref{eq:c0}).
With equation~(\ref{eq:dispersion4}), the instability condition is written as 
\begin{equation}
  -\frac{v_{A}^{2}}{\left(2 - q\right)\Omega} < s\eta_{H} < -\frac{v_{A}^{2}}{2\Omega}.
\end{equation}
As a result, the necessary condition for this instability is $s\eta_{H} < 0$.
In the $\Omega\rightarrow 0$ limit,  the dispersion relation (\ref{eq:dispersion4}) results in ion-cyclotron wave.
Thus, the instability is caused by interaction between ion-cyclotron wave and epicycle motion when $\Omega \ne0$. 
This instability is called `diffusive MRI' in \citet{2012MNRAS.423..222P}. 
We also describe it as `diffusive MRI'  in the following.

\subsection{Summary of Disk Model and Parameters} 
According to the prescription described in \citet{2019MNRAS.484.2119K}, we introduced  a model of the disk  around the protostar, in which the disk is regulated by gravitational instability or  the Toomre parameter $Q$. 
We change the magnetic field strength, magnetic field direction and dust model when constructing the disk. 
We adopt two different strengths of the magnetic field $B=0.1$ and 0.01\,G.
We change the direction of the magnetic field, i.e., parallel ($s=1$) or anti-antiparallel ($s=-1$) to the $z$-direction.  
We also use four different dust models (d0035, d01, MRN, tr-MRN). 
In total, we prepare 16 disk models, as listed in Table~\ref{tab:model_para}. 

For these models, we calculate the chemical reaction to derive the abundance of each species and the electric conductivities.
With the derived conductivities, we calculate the diffusion coefficients of the magnetic  field such as $\eta_O$, $\eta_A$ and $\eta_H$. 
Finally, we evaluate the growth of MRI for these disk models in the range $r=0.1$--$100$\,au and  $z<5h$.

\begin{table}
  \caption{
 Models and parameters.
  }       
  \begin{tabular}{cccc}
  \hline
  Model       & s   & B $(G)$    & Dust  model \\ \hline
  s1B1d0035   & 1   & $0.1$      & d0035 \\ 
  s1B1d01     & 1   & $0.1$      & d01 \\
  s1B1MRN     & 1   & $0.1$      & MRN\\ 
  s1B1trMRN   & 1   & $0.1$      & trMRN \\ 
  s1B2d0035   & 1   & $0.01$     & d0035\\
  s1B2d01     & 1   & $0.01$     & d01 \\ 
  s1B2MRN     & 1   & $0.01$     & MRN \\ 
  s1B2trMRN   & 1   & $0.01$     & trMRN \\ 

  s-1B1d0035  & -1  & $0.1$      & d0035\\ 
  s-1B1d01    & -1  & $0.1$      & d01 \\
  s-1B1MRN    & -1  & $0.1$      & MRN \\ 
  s-1B1trMRN  & -1  & $0.1$      & trMRN \\ 
  s-1B2d0035  & -1  & $0.01$     & d0035 \\
  s-1B2d01    & -1  & $0.01$     & d01 \\ 
  s-1B2MRN    & -1  & $0.01$     & MRN \\ 
  s-1B2trMRN  & -1  & $0.01$     & trMRN \\
  \hline
  \end{tabular}
  \label{tab:model_para}
\end{table}

\section{Results}
\label{sec:results}
We need to determine the abundance of charged particles in order to derive the diffusion coefficients $\eta_O$, $\eta_A$, and $\eta_H$. 
Then, we can judge whether MRI can be activated in each area of the disk by introducing the diffusion coefficients into  equation~(\ref{eq_eval:criterion_mri}).
In this section, we describe the calculation results of the chemical reaction (i.e. the abundance of each species) and  the diffusion coefficients and show the growth rate of MRI in our disk models. 

\subsection{Chemical Abundances and Diffusion Coefficients}
\label{sec:chemicalabundance}
According to the methods described in \S\ref{sec:reactions}, we calculated the chemical reactions with four different dust models in each area of the disk. 
We limited the disk within the range $0.1\,{\rm au} \le r \le 100\,{\rm au}$ in the radial direction and within $0 \le z \le 5h$ 
in the vertical direction, where $r$, $z$ and $h$ mean the distance from the central protostar, 
the distance from the equatorial plane ($z=0$) and the scale height, respectively. 

\begin{figure}
    \centering
    \includegraphics[clip, width=150mm]{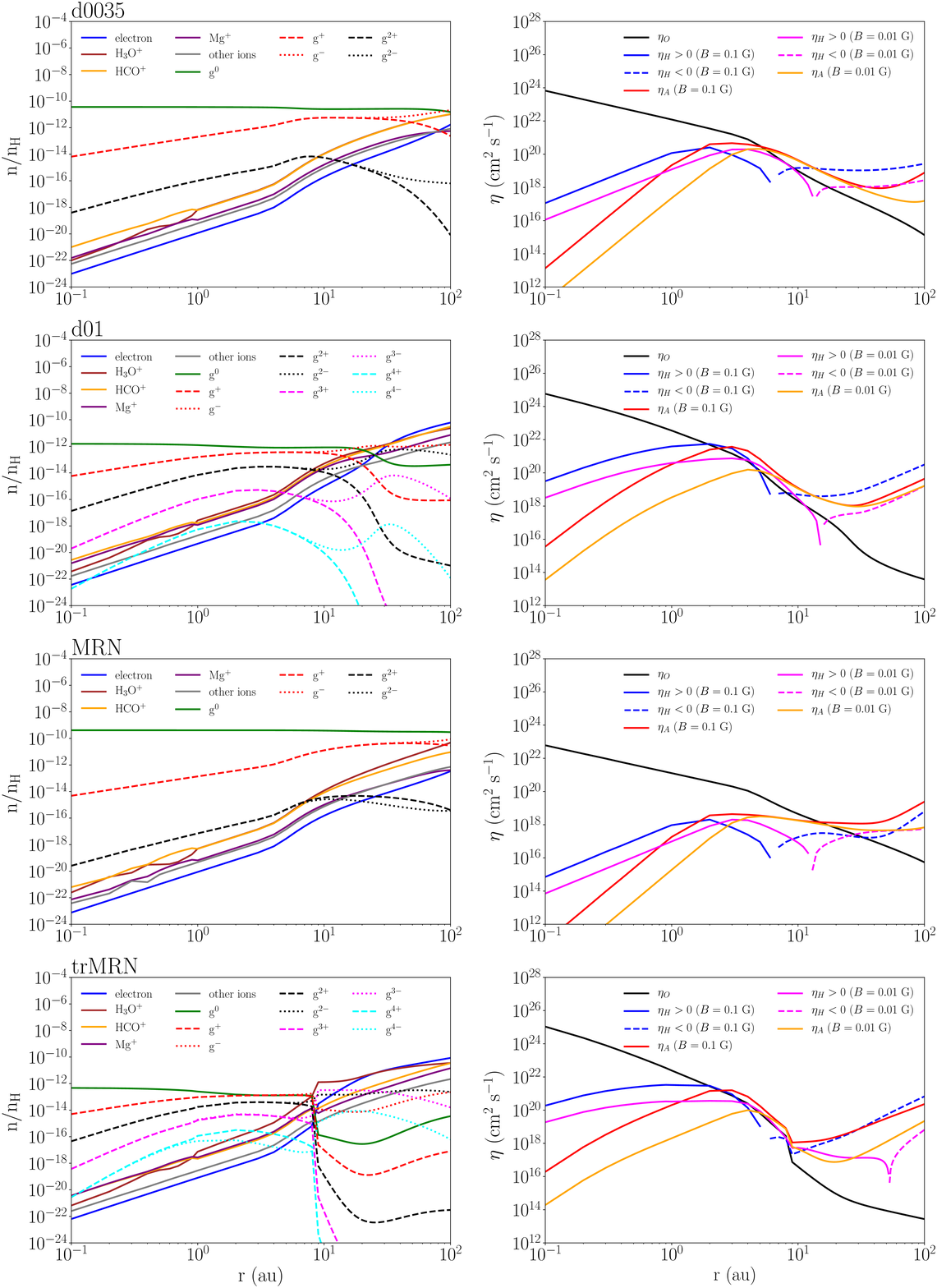}
    \caption{
Abundance of electrons, ions, and dust grains ($\rm{g}^{0}$, $\rm{g}^{\pm 1}$,$\rm{g}^{\pm 3}$, $\rm{g}^{\pm 4}$) to hydrogen atoms $n_H$ (left panels), and the diffusion coefficients  $\eta_O$, $\eta_A$, $\eta_H$ (right panels) on the equatorial plane ($z=0$) as a function of the radius $r$ for the four dust models (indicated in the upper left corner of each left panel). 
Three of the most abundant ions H$_3$O$^+$, HCO$^+$ and Mg$^+$ are also plotted in the left panels, while the sum of any other positively charged atoms and molecules are represented as `other ions.' 
In each right panel, the diffusion coefficients of ambipolar diffusion and the Hall term in the cases of $B=0.01$ and 0.1\,G are plotted, and  the sign of $\eta_H$ is denoted by different line types (solid lines for positive $\eta_H>0$ and broken lines for negative $\eta_H<0$).  
        }
    \label{fig:abund_eta_H0}
\end{figure}

Figure~\ref{fig:abund_eta_H0} plots the chemical abundances (left panels) and the diffusion coefficients (right panels) on the equatorial plane ($z=0$)  against the radius (i.e. distance from the central protostar) for four different dust models. 
In the left panels, although there are differences in the chemical abundances among the models, we can see some similarities. 
The panels show that neutral dust grains ($\rm{g}^{0}$) are most abundant in the inner disk region.
The neutral grains are most abundant over the whole range ($0.1\,{\rm au} < r < 100\,{\rm au}$) for d0035 and MRN.
On the other hand, ions (H$_3$O$^+$, HCO$^+$ and Mg$^+$) and electrons dominate the neutral  grains in the range of $r\gtrsim 20$--$30$\,au for d01 and tr-MRN.

The neutral grains do not directly contribute to the estimation of the conductivities and magnetic diffusivities.
Among the charged particles, singly charged grains ($\rm{g}^{+}$ and $\rm{g}^{-}$) are most abundant in the inner disk region of $r\lesssim10$\,au for all dust models. 
In the outer disk region of $r\gtrsim 10$\,au,  the abundances of ions and electrons are comparable to the singly charged grains for d0035, d01 and MRN and they greatly dominate the singly charged grains for tr-MRN. 
In the region around $r\sim100$\,au, negatively charged grains ($\rm{g}^{-}$) and HCO$^+$ ions are the most abundant for d0035.
On the other hand, electron and ions (H$_3$O$^+$, HCO$^+$ and Mg$^+$) are more abundant than other charged particles at $r\sim100$\,au for d0035 and tr-MRN. 
For MRN, the abundances of charged grains ($\rm{g}^{+}$, $\rm{g}^{-}$) are still comparable to those of electrons and ions at $r\sim100$\,au.
Note that grains with large charge number ($\rm{g}^{2+}$, $\rm{g}^{2-}$, $\rm{g}^{3+}$, $\rm{g}^{3-}$, $\rm{g}^{4+}$, $\rm{g}^{4-}$) 
are always less abundant than singly charged grains ($\rm{g}^{+}$, $\rm{g}^{-}$) over the whole range (0.1--100\,au).  

Small dust grains effectively absorb the charged particles of ions and electrons because of their large surface area.  
Thus, charged particles tend to be less abundant for models with small grains (d0035 and MRN) than for the models with large grains (d01 and tr-MRN). 
As a result, the dust properties influence the chemical abundance of each area of the disk.

The difference in the abundance of charged particles causes a difference in the diffusion coefficients $\eta_O$, $\eta_A$, and $\eta_H$. 
The diffusion coefficients on the equatorial plane are plotted against radius in Figure~\ref{fig:abund_eta_H0} right panels.
Since $\eta_A$ and $\eta_H$ depend on the magnetic field strength $B$, we plotted them for two cases of $B=0.1$ and 0.01\,G.
Note that we adopted a uniform distribution of the magnetic field as described in \S\ref{sec:magneticfield}.
In addition, the diffusion coefficient $\eta_H$ has both negative and positive signs, which are described by different line types  in the panels.

As seen in Figure~\ref{fig:abund_eta_H0} right panels, there are some similarities  in the distribution of the coefficients among all the models. 
$\eta_{\rm O}$ dominates both $\eta_{\rm A}$ and $\eta_{\rm H}$ in the inner disk region of $r\lesssim 1$--$10$\,au. 
$\eta_{\rm A}$ and $\eta_{\rm H}$ gradually increase as the distance from the protostar increases, and they dominate $\eta_O$ at $r \sim 1$--$10$\,au. 
$\eta_{\rm A}$ and $\eta_{\rm H}$ dominate $\eta_{\rm O}$ in the outer disk region of $r\gtrsim 1$--$10$\,au. 

The differences in the coefficients between the models are noticeable in the outer disk region. 
A negative $\eta_{\rm H}$ ($\eta_{\rm H}<0$) is largest in the outer disk region  ($r \sim100$\,au) for  d0035,  d01 and tr-MRN when the magnetic field strength is $B=0.1$\,G. 
On the other hand, $\eta_{\rm A}$ dominates $\eta_{\rm O}$ and $\eta_{\rm H}$ at $r\sim100$\,au for MRN. 
The three coefficients ($\eta_{\rm O}$, $\eta_{\rm A}$ and $\eta_{\rm H}$) are comparable in the  intermediate region ($r\sim1$--$10$\,au).
$\eta_{\rm A}$ and $\eta_{\rm H}$ are smaller for $B=0.01$\,G than for $B=0.1$\,G. 
$\eta_{\rm A}$ and $\eta_{\rm H}$ also become small as the magnetic field strength weakens. 
On the other hand,  $\eta_{\rm O}$ does not depend on the magnetic field strength, as descried in \S\ref{sec:nonidealMHD}.
Meanwhile, $\eta_{\rm O}$ monotonically decreases as the radius increases in all models.
Thus, the difference between $\eta_{\rm O}$ and the other two, $\eta_{\rm A}$ and $\eta_{\rm H}$, is sensitive to the magnetic field strength.

\begin{figure}
    \centering
    \includegraphics[clip, width=150mm]{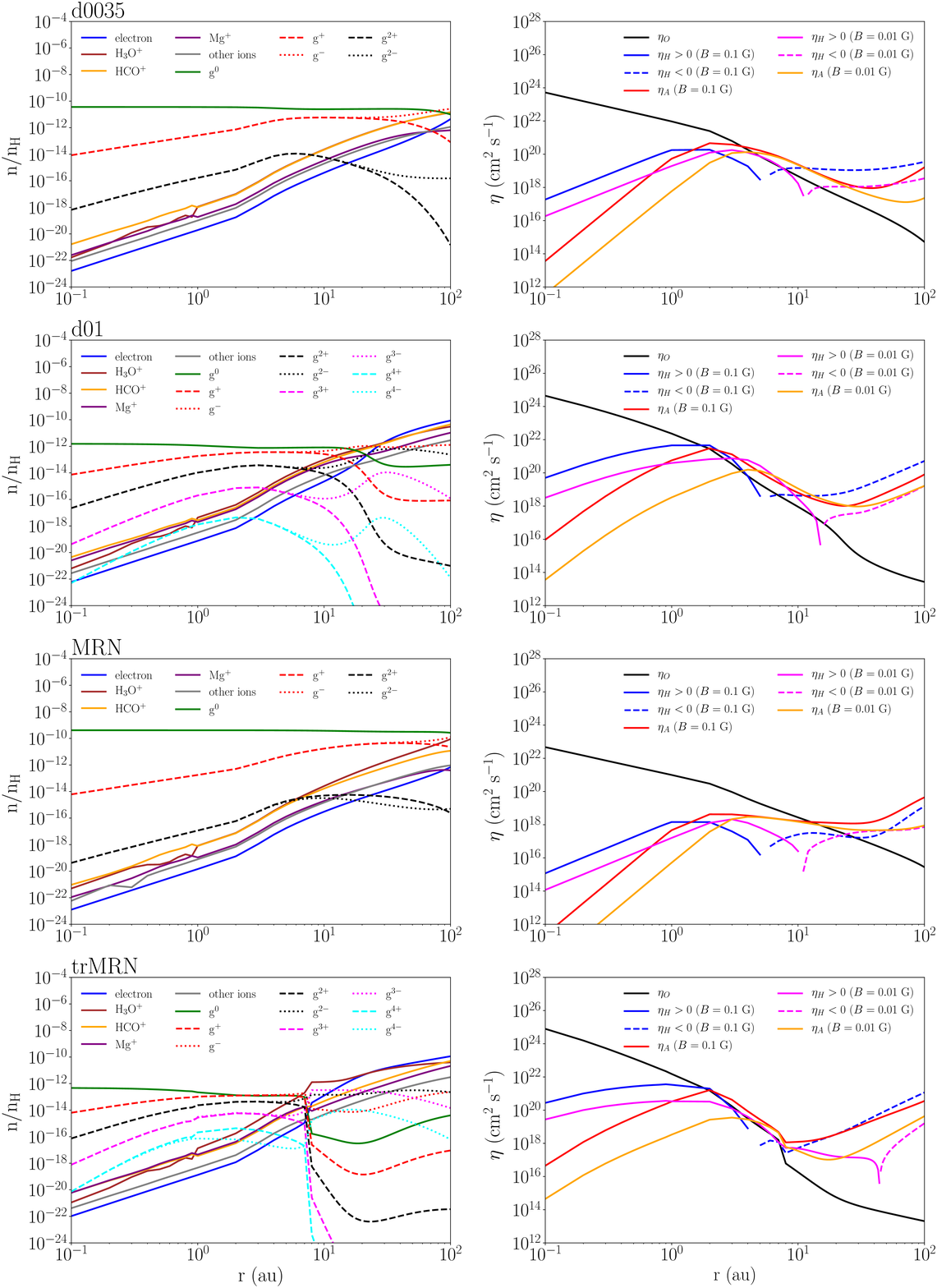}
    \caption{
        As Fig.~\ref{fig:abund_eta_H0}, but on the $z=h$ plane. 
        }
    \label{fig:abund_eta_H1}
\end{figure}

The abundance of each type of species and the diffusion coefficients on the $z=h$ plane are plotted in Figure~\ref{fig:abund_eta_H1}.
Since we assume a local isothermal and uniform magnetic field, the temperature and magnetic field strength on the $z=h$ plane are the same as those on the $z=0$ plane.  
Meanwhile, the density on the $z=h$ plane is different from that on the equatorial plane: the density on the $z=h$ plane is 0.606 ($\exp(-1/2)$) times lower than that on the $z=0$ plane.  
Figure~\ref{fig:abund_eta_H1} shares many similarities to Figure~\ref{fig:abund_eta_H0}, although they do have minor differences. 
As seen in the left panels of Figure~\ref{fig:abund_eta_H1}, singly charged dust grains are the most abundant in the inner disk region, 
while electrons and ions are the most abundant species around $r\sim100$\,au, which is the same as in Figure~\ref{fig:abund_eta_H0} left panels. 
Since the abundance distributions on the $z=h$ plane are almost the same as those on the $z=0$ plane, there is no significant difference 
in the diffusivities  (Fig.~\ref{fig:abund_eta_H1} right panels) between them.

\begin{figure}
    \centering
    \includegraphics[clip, width=150mm]{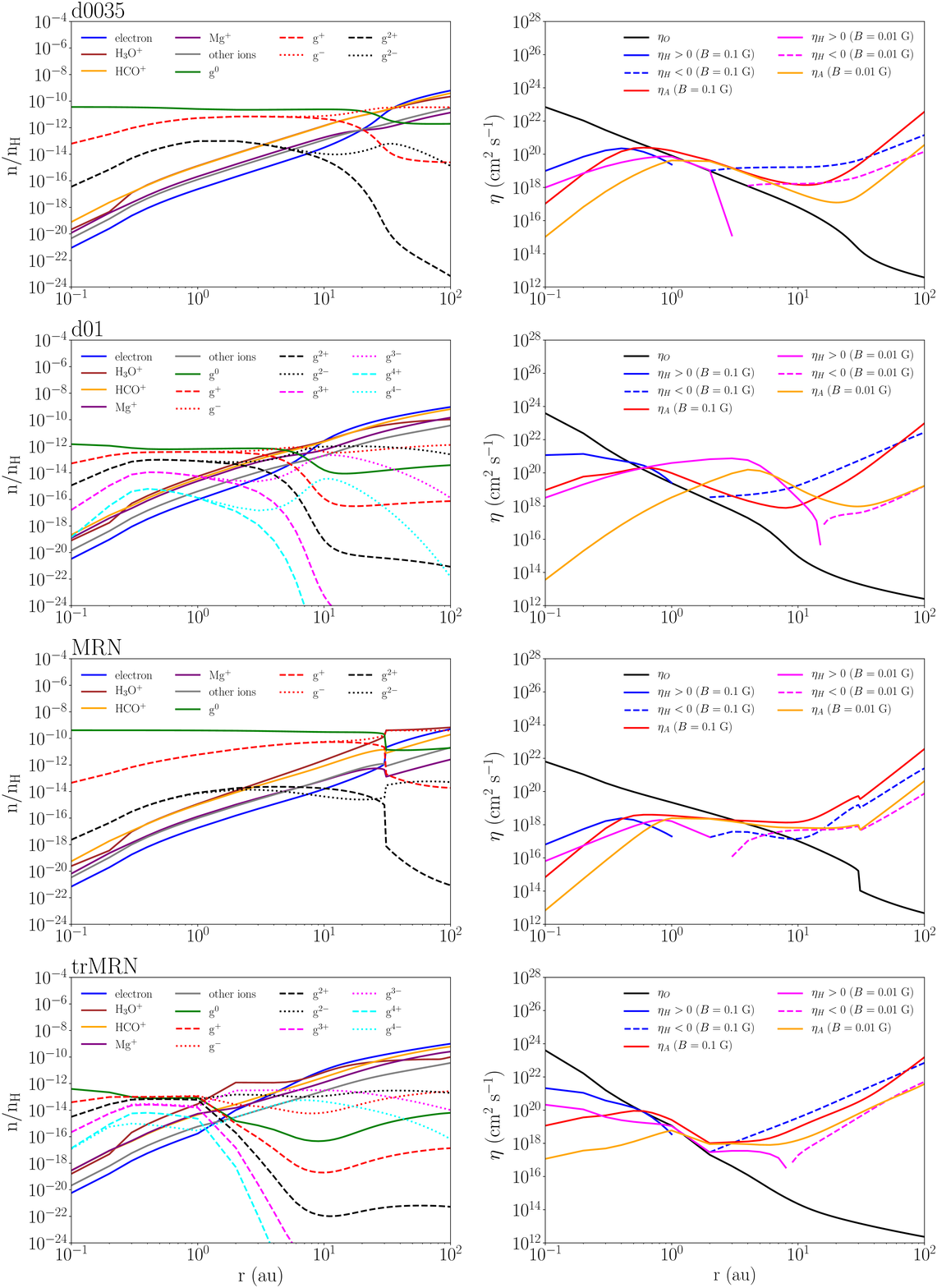}
    \caption{
        As Fig.~\ref{fig:abund_eta_H0}, but on the $z=3h$ plane. 
        }
    \label{fig:abund_eta_H3}
\end{figure}

Figure~\ref{fig:abund_eta_H3} shows the abundance distributions (left panels) and magnetic diffusivities (right panels) on the $z=3h$ plane. 
Although the distributions of chemical abundances on the $z=3h$ plane are qualitatively the same as those on the $z=0$ plane, there are quantitative differences.
Compared with the abundances on the equatorial (or $z=0$ plane, Fig.~\ref{fig:abund_eta_H1} left panels),  
electrons and ions are relatively abundant on the $z=3h$ plane (Fig.~\ref{fig:abund_eta_H3} left panels). 
The density on the $z=3h$ plane is 0.011 times ($\exp(-9/2)$) lower than that on the $z=0$ plane. 
The collisional rate  of electrons and ions onto dust grains is low  because of a relatively low density on the $z=3h$ plane. 
Therefore, the ionization degree on the $z=3h$ plane is higher than that on the equatorial plane, 
which results in relatively small diffusion coefficients as seen in Figure~\ref{fig:abund_eta_H3} right panels.

Meanwhile,  the tendency of the distributions of the diffusion coefficients for $z=3h$ is almost the same as on the $z=0$ plane. 
$\eta_{\rm O}$ is much larger than $\eta_{\rm A}$ and $\eta_{\rm H}$ in the inner disk region of $r\lesssim 1$--$10$\,au, while $\eta_{\rm A}$ and $\eta_{\rm H}$ are largest in the range of $r\gtrsim 10$\,au. 
Quantitative differences in the diffusion coefficients should affect the growth rate of MRI, as explained in the following sections. 

Figures~\ref{fig:abund_eta_H0}--\ref{fig:abund_eta_H3} indicate that the abundance of each species and diffusion coefficients strongly depend on 
both the dust model and the position of the circumstellar disk. 
Using the diffusion coefficients plotted in these figures, we determine whether MRI occurs, as detailed below.

\subsection{Comparison among Diffusion Coefficients}
\begin{figure}
    \centering
    \includegraphics[clip, width=140mm]{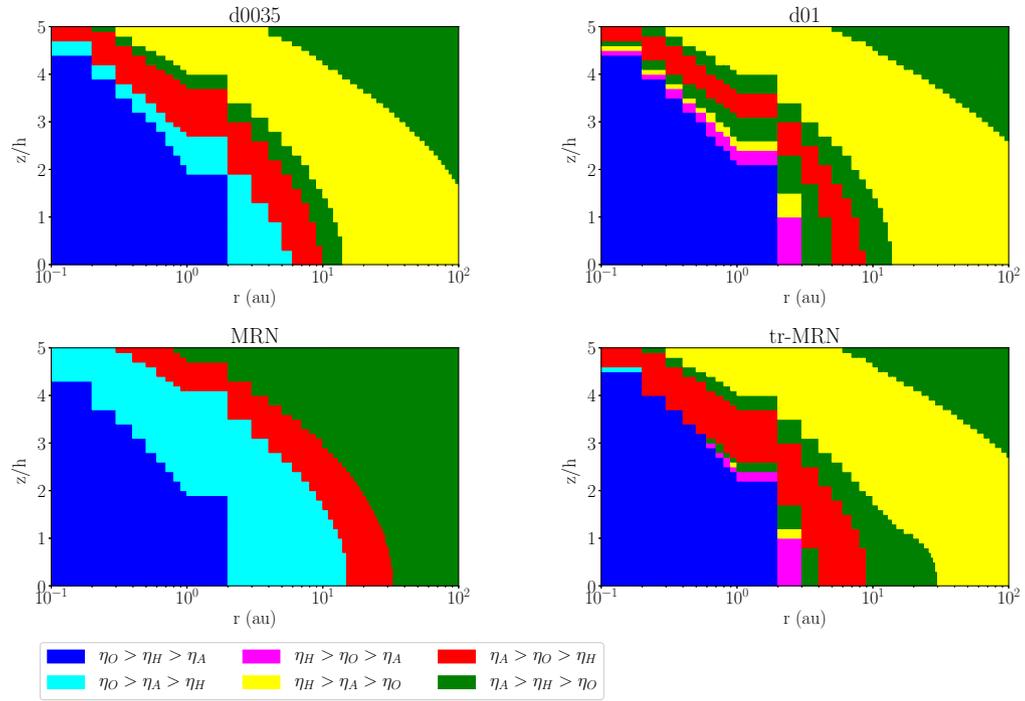}
    \caption{
Magnitude relation of resistivities $\eta_{\rm O}$, $\eta_{\rm A}$ and $\eta_{\rm H}$ on the $r$--$z$ plane for four different dust models. 
A magnetic field strength of $B=0.1$\,G is assumed to estimate the resistivities. 
The dust model is indicated above each panel. 
}
\label{fig:resisB01}
\end{figure}

We show the diffusion coefficients of the magnetic field  only on the $z=0$, $h$, $3h$ planes in \S\ref{sec:chemicalabundance}. 
In this subsection, we compare these three coefficients ($\eta_{\rm O}$, $\eta_{\rm A}$, $\eta_{\rm H}$)  for the whole region of the disk. 
The magnitude relation of the resistivities or diffusion coefficients with $B=0.1$\,G is plotted on the $r$--$z$ plane in Figure~\ref{fig:resisB01}. 
It should be noted that the magnitude relation is not closely related to the growth of MRI and the amplitude of each coefficient 
(not magnitude relation) is significant for determining the growth condition of MRI. 
Here, we aim to understand the condition and state of our disk model with comparison among the resistivities  (or diffusion coefficients).

For the single dust model d0035 (Fig.~\ref{fig:resisB01} top left), $\eta_{\rm O}$ is largest in the inner disk region, 
while $\eta_{\rm A}$ dominates both $\eta_{\rm O}$ and $\eta_{\rm H}$ in the disk intermediate region outside which $\eta_{\rm H}$ becomes the largest.
Although other models have almost the same tendency as model d0035, there are some differences.
For  the d01 and tr-MRN dust models, which contain relatively large-sized grains, $\eta_{\rm H}$ 
dominates the other two coefficients in the range $2\lesssim r \lesssim 3$\,au on the equatorial plane. 
For the MRN model, $\eta_{\rm H}$ never dominates both $\eta_{\rm O}$ and $\eta_{\rm A}$, as seen in Figure~\ref{fig:resisB01} bottom left panel.
For this model, $\eta_{\rm O}$ dominates $\eta_{\rm A}$ and $\eta_{\rm H}$ in the inner disk region, while $\eta_{\rm A}$ 
dominates $\eta_{\rm O}$ and $\eta_{\rm H}$ in the outer disk region. 

\begin{figure}
    \centering
    \includegraphics[clip, width=140mm]{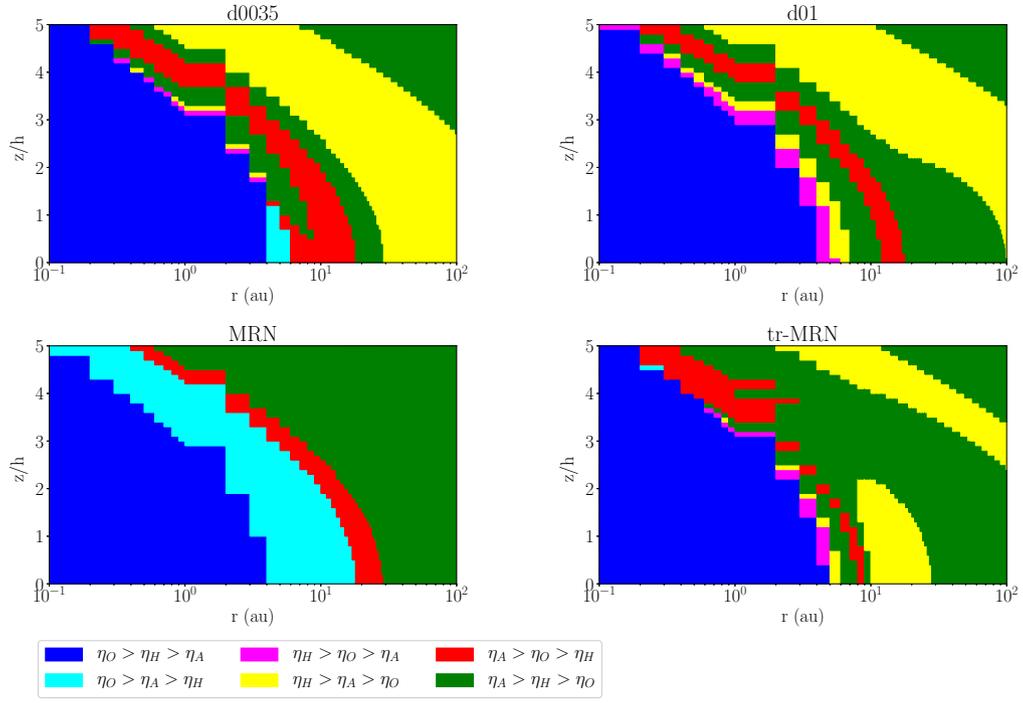}
    \caption{
As Fig.~\ref{fig:resisB01}, but with a magnetic field strength of $B=0.01$\,G adopted to estimate the resistivities. 
}
\label{fig:resisB001}
\end{figure}

The magnitude relation of the diffusivities with a weak magnetic field $B=0.01$\,G is plotted in Figure~\ref{fig:resisB001}. 
The trend of the magnitude relation is not significantly changed with a weak magnetic field. 
However, both $\eta_{\rm A}$ and $\eta_{\rm H}$ decrease as the magnetic field weakens, while $\eta_{\rm O}$ does not depend on the magnetic field strength.  
As a result, the region where $\eta_{\rm O}$ dominates widens compared with the strong magnetic field case ($B=0.1$\,G, Fig.~\ref{fig:resisB001}). 
In addition, $\eta_{\rm A}$ tends to dominate $\eta_{\rm H}$ especially in the outer disk region. 
For example, for d01 and tr-MRN, the region where $\eta_{\rm H}$ dominates is narrower in the weak magnetic field models (Fig.~\ref{fig:resisB001}) 
than in the strong magnetic field models (Fig.~\ref{fig:resisB01}).

\subsection{MRI Growth in Disk}
\subsubsection{Strong Magnetic Field Parallel to Angular Momentum}
\label{sec:s1}
\begin{figure}
    \centering
    \includegraphics[clip, width=160mm]{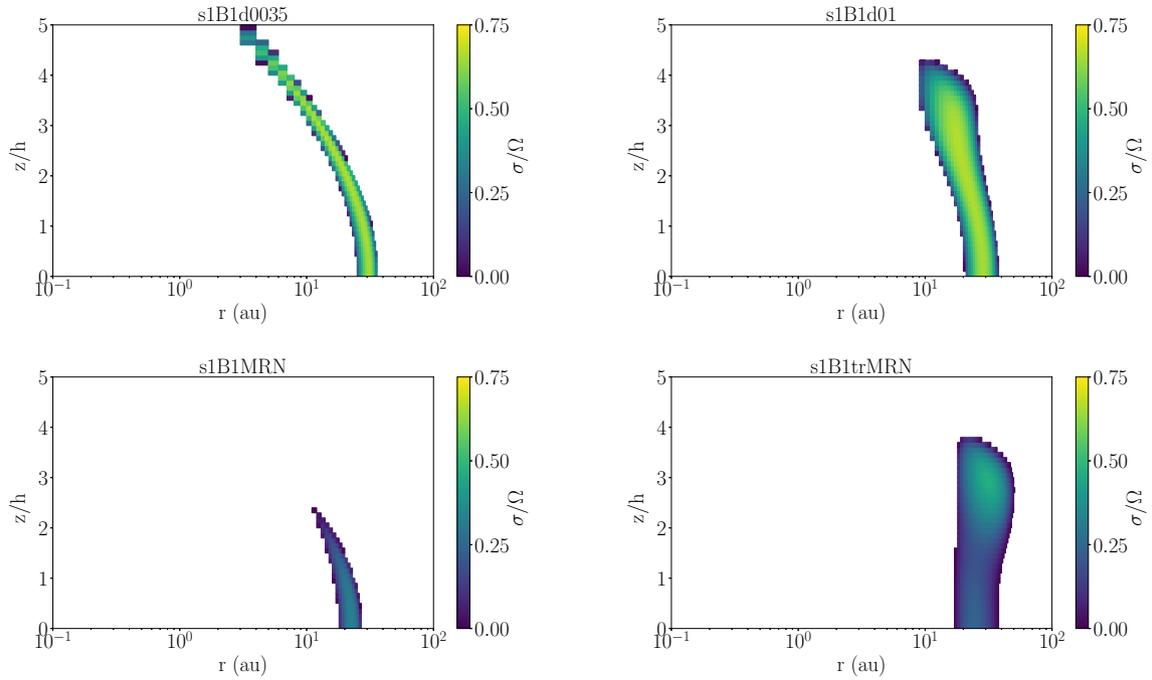}
    \caption{
            Growth rate of MRI on the $r$--$z$ plane for models s1B1d0035 (top left), s1B1d01 (top right), s1B1MRN (bottom left) and s1B1trMRN (bottom right). 
            The models have the same parameters of  $B=0.1$\,G and $s=1$, but are different dust models. 
            The $z$-direction is normalized by the scale height at each radius $r$. 
            The color denotes the growth rate of the fastest growth mode normalized by the Keplerian angular velocity. 
            }
    \label{fig:B1-s1}
\end{figure}

We show the growth rate of MRI in our disk model  when the magnetic field vector is parallel to the rotation vector ($s=1$). 
Figure~\ref{fig:B1-s1} plots  the growth rate of MRI normalized by the Keplerian angular velocity on the $r$--$z$ plane for four models s1B1d0035 (top left), 
s1B1d01 (top right), s1B1MRN (bottom left) and s1B1trMRN (bottom right). 
The parameters of the disk model are described in Table~\ref{tab:model_para}.
In each panel of Figure~\ref{fig:B1-s1}, the colored area corresponds to the MRI unstable region 
where the growth rate of the fastest growth mode normalized by the Keplerian angular velocity is denoted by the color. 
On the other hand, the white region (i.e. uncolored  region) is stable against the MRI, indicating that MRI does not grow in this region. 
The models shown in Figure~\ref{fig:B1-s1} have the same magnetic field strength $B=0.1$\,G, but are different dust models. 
Thus, the difference in the MRI active region and growth rate can be attributed to the dust model.
The maximum growth rate for both the single dust models s1B1d0035 (top left) and s1B1d01 (top right) is $\sigma/\Omega \simeq 0.75$, 
which indicates that the MRI can develop in the colored region within the Keplerian timescale.
The growth rate of the dust size-distribution models s1B1MRN and s1B1trMRN is slightly smaller than 
that in s1B1d0035 and s1B1d01, while the distribution of the MRI active region (colored region) is almost the same between the models. 

In all the models shown in Figure~\ref{fig:B1-s1}, roughly speaking, the MRI growth is suppressed by ohmic dissipation in the inner disk region.
In more detail, although the diffusion coefficient of ohmic dissipation dominates other two in the inner disk region (Fig.~\ref{fig:resisB01}), both ambipolar diffusion and Hall effect cannot be completely ignored for the suppression of MRI there. 
On the other hand,  the MRI growth is mainly suppressed by ambipolar diffusion in the outer disk region. 
As seen in Figure~\ref{fig:resisB01}, either $\eta_{H}$ or $\eta_{A}$ dominates $\eta_{O}$ in the range  $r\gtrsim10-20$\,au. 
Among the dust distribution models MRN and tr-MRN, $\eta_{A}$ dominates (or is comparable to) $\eta_{H}$. 
In such a case ($\eta_{A}$ being comparable to $\eta_{H}$), the  ambipolar diffusion plays almost the same role as the ohmic dissipation when only considering the vertical component of  the magnetic field (i.e. $B_z$ component), in which small-scale ($k<h$) unstable modes are dumped by  ambipolar diffusion.
Thus, in the outer disk region, the MRI growth is effectively suppressed by ambipolar diffusion especially in the dust distribution models s1B1MRN and s1B1trMRN.
It should be noted that, in addition to ambipolar diffusion and Hall effect, the MRI growth can be also suppressed due to the strong magnetic field in the range  $r\gtrsim 40$\,au. 
The contribution of ohmic dissipation, ambipolar diffusion and Hall effect and the effect of magnetic field strength will be precisely presented in a subsequent paper.
 
\begin{figure}
    \centering
    \includegraphics[clip, width=145mm]{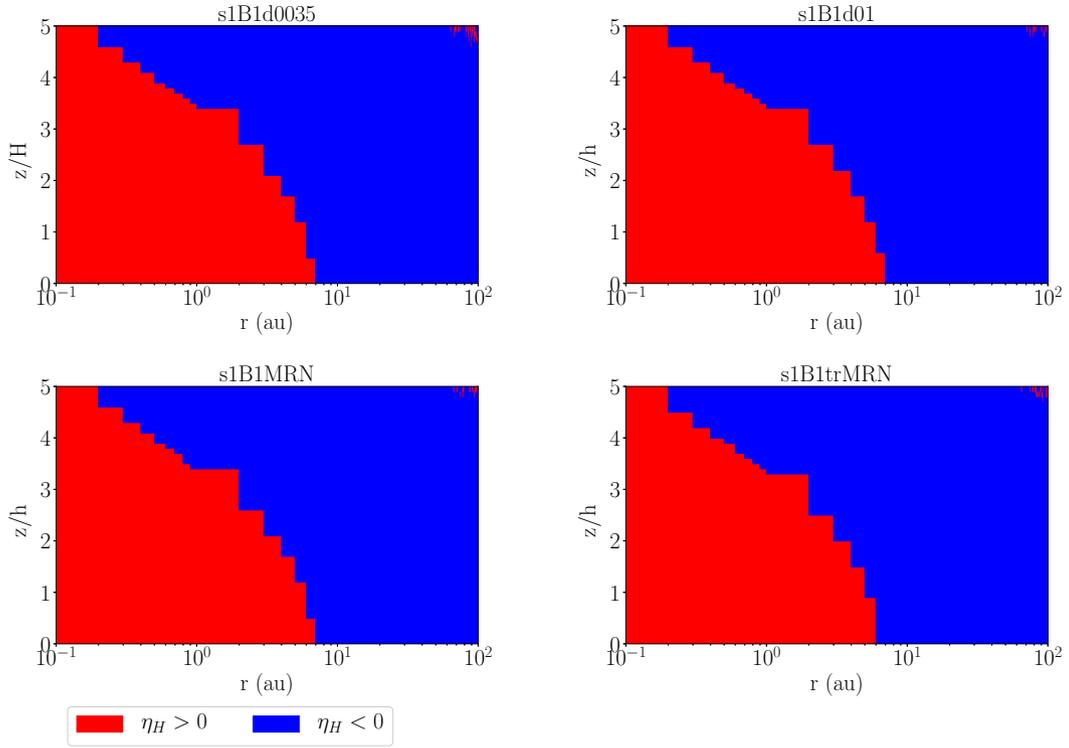}
    \caption{
    Distribution of the sign of the Hall coefficient $\eta_{H}$ on the $r$--$z$ plane for models s1B1d0035 (top left), s1B1d01 (top right), s1B1MRN (bottom left) and s1B1trMRN (bottom right). 
    The Hall coefficient is positive ($\eta_{H}>0$) in the red colored region, while it is negative ($\eta_{H}<0$) in the blue colored region. 
    }
    \label{fig:B1-etaHsgn}
\end{figure}

Next, we focus  on the MRI active region (colored region). 
The MRI active region (or colored region) distributed in the range $20$--$30$\,au is considered to be due to the Hall effect. 
As described in \S\ref{sec:crit_mri}, the Hall effect can cause the HSI and diffusive MRI, in which the sign of $s\eta_{H}$ determines which instability grows. 
The HSI  develops when $s\eta_{H}>0$, while the diffusive MRI grows when $s\eta_{H}<0$. 
The sign of $\eta_H$ is plotted in  Figure~\ref{fig:B1-etaHsgn}.

In all the models, the sign of $\eta_{H}$ is positive  in the range $r\lesssim 7$\,au on the equatorial plane. 
All the MRI active areas (colored area) in Figure~\ref{fig:B1-s1} are covered by the negative $\eta_H$  regions (or blue regions) in  Figure~\ref{fig:B1-etaHsgn}.
This indicates that diffusive MRI drives MRI \citep{2012MNRAS.422.2737W,2012MNRAS.423..222P} because of $s\eta_{H}<0$.
In addition, $\eta_{\rm H}$ dominates both $\eta_{\rm O}$ and $\eta_{\rm A}$ in the colored region of Figure~\ref{fig:B1-s1}, except for model s1B1MRN (see Fig.~\ref{fig:resisB01}). 
Note that the Hall coefficient cannot be the largest for the whole disk area for the MRN model in Figure~\ref{fig:resisB01}. 
Thus,  the MRI active region is the smallest in the MRN model (model s1B1MRN) among the models shown in Figure~\ref{fig:B1-s1}. 
Therefore, the MRI active region is attributed to the Hall effect.

\subsubsection{Weak Magnetic Field Parallel to Angular Momentum}
\begin{figure}
    \centering
    \includegraphics[clip, width=160mm]{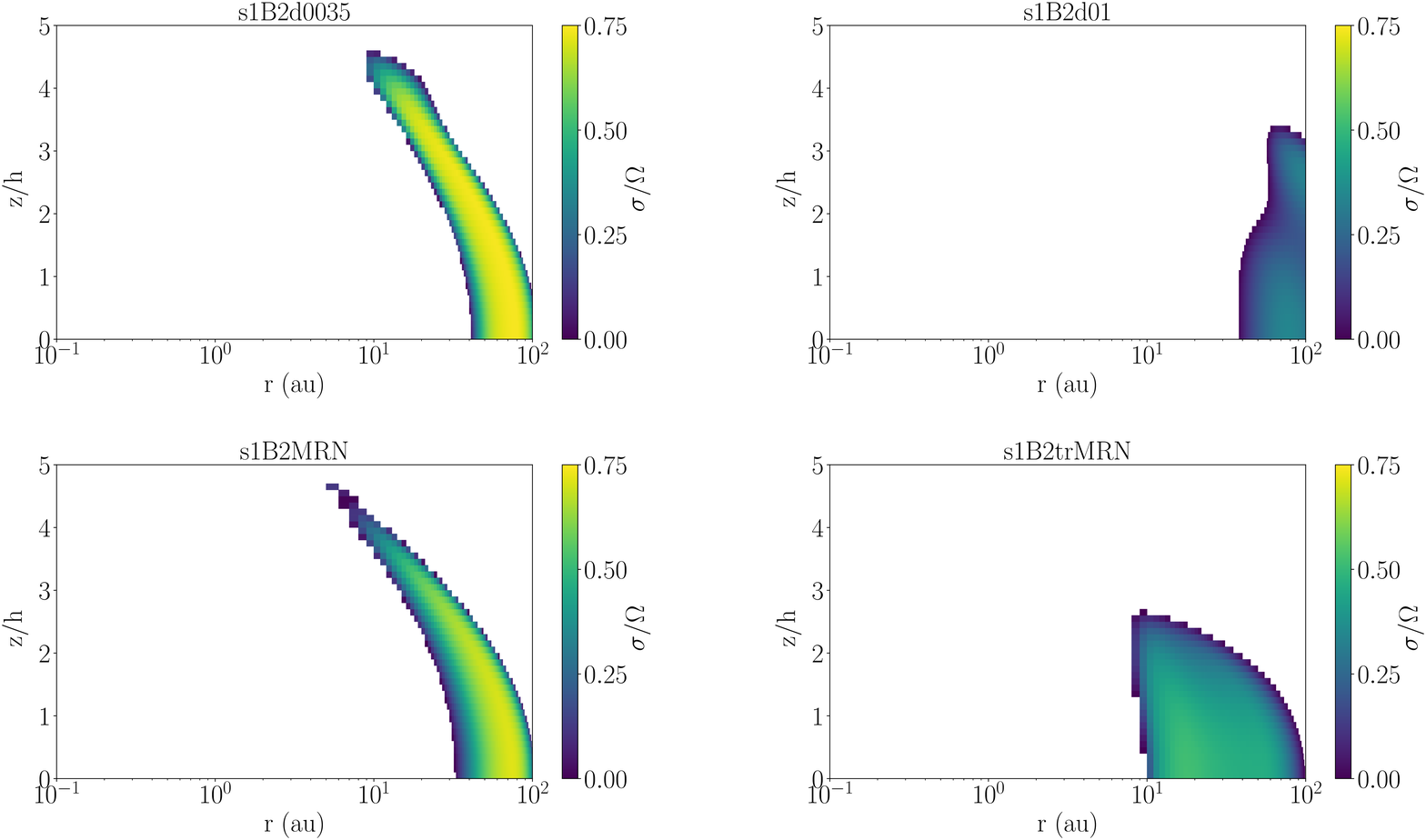}
    \caption{           
            As Fig.~\ref{fig:B1-s1} but for the models s1B2d0035 (top left), s1B2d01 (top right), s1B2MRN (bottom left), s1B2trMRN (bottom right). 
            The models have the same parameters of  $B=0.01$\,G and $s=1$, but are different dust models. 
            }
    \label{fig:B2-s1}
\end{figure}

Next, we focus on the models with a weak magnetic field $B=0.01$\,G when  the magnetic field is  parallel to the rotation vector of the disk ($s$ = $1$). 
The MRI growth rate on the $r$--$z$ plane for the models with $B=0.01$\,G and $s=1$ (s1B2d0035, s1B2d01, s1B2MRN and s1B2trMRN) is plotted in Figure~\ref{fig:B2-s1}. 
In the figure, the MRI active area is wider  in the models with $B=0.01$\,G (weak magnetic field models, Fig.~\ref{fig:B2-s1}) than in the models with $B=0.1$\,G 
(strong  magnetic field models, Fig.~\ref{fig:B1-s1}). 
Since MRI tends to occur with a weak magnetic field, it is natural that it will occur in a large area  in Figure~\ref{fig:B2-s1}.

We can confirm a similar distribution of the MRI active region in models s1B2d0035 (Fig.~\ref{fig:B2-s1} top left) and s1B2MRN (Fig.~\ref{fig:B2-s1} bottom left). 
s1B2d0035 has a single grain size of $a = 0.035\,\mu$m, while s1B2MRN has an MRN size distribution. 
For these models,  MRI can grow in the range $r\gtrsim 30$--$40$\,au on the equatorial plane ($z=0$).
Thus, MRI should occur  in the outer disk region. 
The MRI active region is also distributed in the range $r\gtrsim 40$\,au for the model with a single grain size of $a = 0.1\,\mu$m (model s1B2d01, Fig.~\ref{fig:B2-s1} top right).
The MRI active area for the model with tr-MRN distribution (model s1B2trMRN) is the widest on the equatorial plane ($z$ = $0$) among the models shown in Figure~\ref{fig:B2-s1}. 
Although the MRI active region far from (or above) the equatorial plane depends on the radius $r$, it is distributed in the range $\gtrsim 5\,{\rm au}$.

The growth rate of MRI for model s1B2trMRN is lower than that for s1B2d0035 and s1B2MRN, while it is still as high as $\sigma/\Omega \sim 0.5$. 
Thus, MRI could grow in the colored area  within the Keplerian timescale for s1B2trMRN. 
In addition, the growth rate of MRI is comparable to the Keplerian timescale for s1B2d0035 and s1B2MRN. 
Thus, MRI should activate in a short timescale in the colored region for these models. 
On the other hand,  the growth rate of MRI is $\sigma/\Omega \lesssim 0.2$ in the colored area for s1B2d01.
Thus, it is not clear that MRI can sufficiently grow for this model.
It should be noted that the growth rate of MRI for s1B2d01 (Fig.~\ref{fig:B2-s1} top right)  is considerably lower than that for s1B2d0035 (Fig.~\ref{fig:B2-s1} top left). 
A single dust size is adopted for both s1B2d01 ($a_{\rm d}=0.1\mu$\,m)  and s1B2d0035 ($a_{\rm d}=0.035\mu$\,m). 
Thus, a slight difference in the dust size causes a significant difference in the MRI growth rate.

\begin{figure}
    \centering
    \includegraphics[clip, width=145mm]{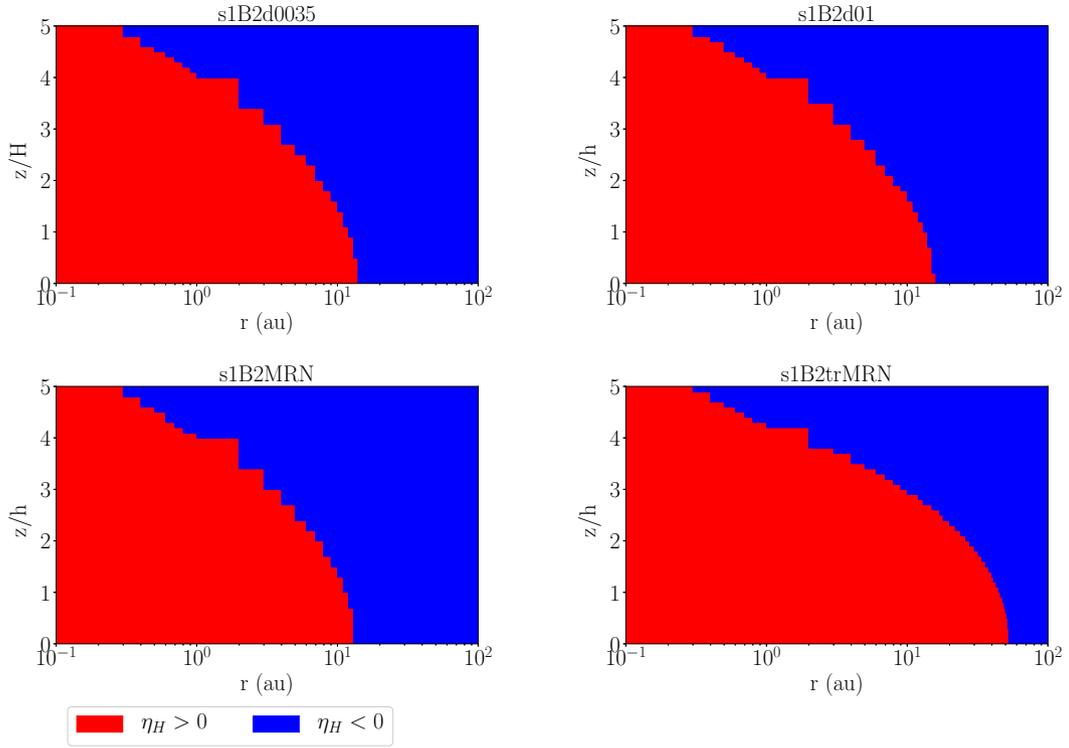}
    \caption{
        As Fig.~\ref{fig:B1-etaHsgn} but for different models s1B2d0035 (top left), s1B2d01 (top right), s1B2MRN (bottom left), s1B2trMRN (bottom right).  
    }
    \label{fig:B2-etaHsgn}
\end{figure}

In addition to the standard MRI that  is induced without non-ideal MHD terms, the Hall effect also contributes to the MRI growth in the MRI active region.
The sign of the Hall coefficient  for the same models of Figure~\ref{fig:B2-s1} is plotted in Figure~\ref{fig:B2-etaHsgn}. 
In the figure, a positive Hall coefficient $\eta_{H}>0$ corresponding to the red colored area is distributed in the range $r<15$\,au for s1B2d0035, 
s1B2d01 and s1B2MRN, within which the ohmic dissipation or ambipolar diffusion coefficient dominates the Hall coefficient (Fig.~\ref{fig:resisB001}). 
Thus, either ohmic dissipation or ambipolar diffusion suppresses the growth of MRI, and the HSI is not efficient enough to activate MRI. 
In other words, in this region, although HSI is induced with $s\eta_{H}>0$, both ohmic dissipation and ambipolar diffusion do not enhance the  MRI growth. 

On the other hand, unlike these models, the positive  $\eta_{H}>0$ area is distributed in the range $r\lesssim 50$\,au for the model with a tr-MRN distribution (s1B2trMRN). 
Thus, for s1B2trMRN, it is considered that an unstable mode is caused by HSI because the MRI active area (Fig.~\ref{fig:B2-s1} bottom right) is partly superimposed onto 
the positive $\eta_H$ ($\eta_H>0$) area  (Fig.~\ref{fig:B2-etaHsgn} bottom right). 
In addition, the Hall coefficient in this area  is the first or second largest among the three diffusivities (see Fig.~\ref{fig:resisB001}).
As a result, a wide MRI active area  is realized for s1B2trMRN, as seen in the bottom right panel of Figure~\ref{fig:B2-s1}. 
Meanwhile, diffusive MRI contributes to the growth of MRI in the region of $\eta_{H}<0$ or $s\eta_{H}<0$ (see \S\ref{sec:strongBs-1}). 

\subsubsection{Strong Magnetic Field Anti-parallel to Angular Momentum}
\label{sec:strongBs-1}
Here, we show the MRI growth rate when the magnetic field  is anti-parallel to the angular momentum vector in the disk.  
As described in \S\ref{sec:s1}, the  HSI, which is induced by the Hall effect, promotes the growth of MRI, 
in which  the Hall term is affected by the sign (or direction) of the magnetic field as described in \ref{sec:crit_mri}.

\begin{figure}
    \centering
    \includegraphics[clip, width=160mm]{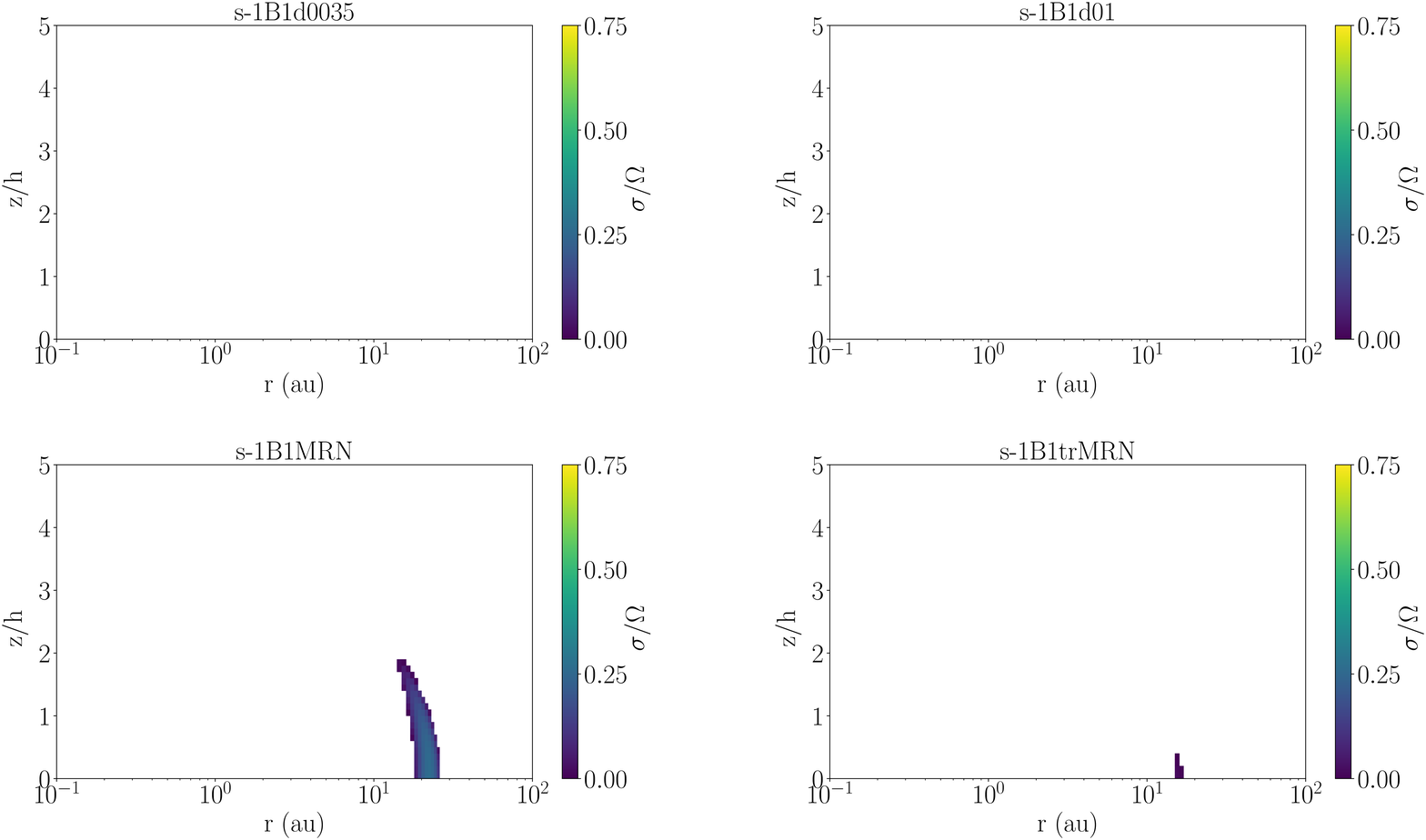}
    \caption{           
            As Fig.~\ref{fig:B1-s1} but for different models s-1B1d0035 (top left), s-1B1d01 (top right), s-1B1MRN (bottom left), s-1B1trMRN  (bottom right). 
            The models have the same parameters of  $B=0.1$\,G and $s=-1$, but are different dust models. 
            }
    \label{fig:B1-s-1}
\end{figure}

The MRI growth rate for the models with $B=0.1$\,G and $s=-1$ (magnetic vector anti-parallel to angular momentum vector) is plotted in Figure~\ref{fig:B1-s-1}. 
As seen in the figure, MRI is significantly suppressed in almost all the areas of the disk. 
Although a very small active area of MRI exists in  s-1B1MRN and s-B1trMRN, the growth rate of MRI $\sigma/\Omega$ is very small.
Thus, it is difficult for MRI to grow even in these areas.  
For these models, the Hall coefficient is negative $\eta_{H} < 0$ in a large part of the disk as shown in Figure~\ref{fig:B1-etaHsgn} and $s\eta_{H} > 0$.
Thus, diffusive MRI contributes to the stabilization of MRI. 

\subsubsection{Weak Magnetic Field Anti-parallel to Angular Momentum}
\label{sec:weakBs-1}
\begin{figure}
    \centering
    \includegraphics[clip, width=160mm]{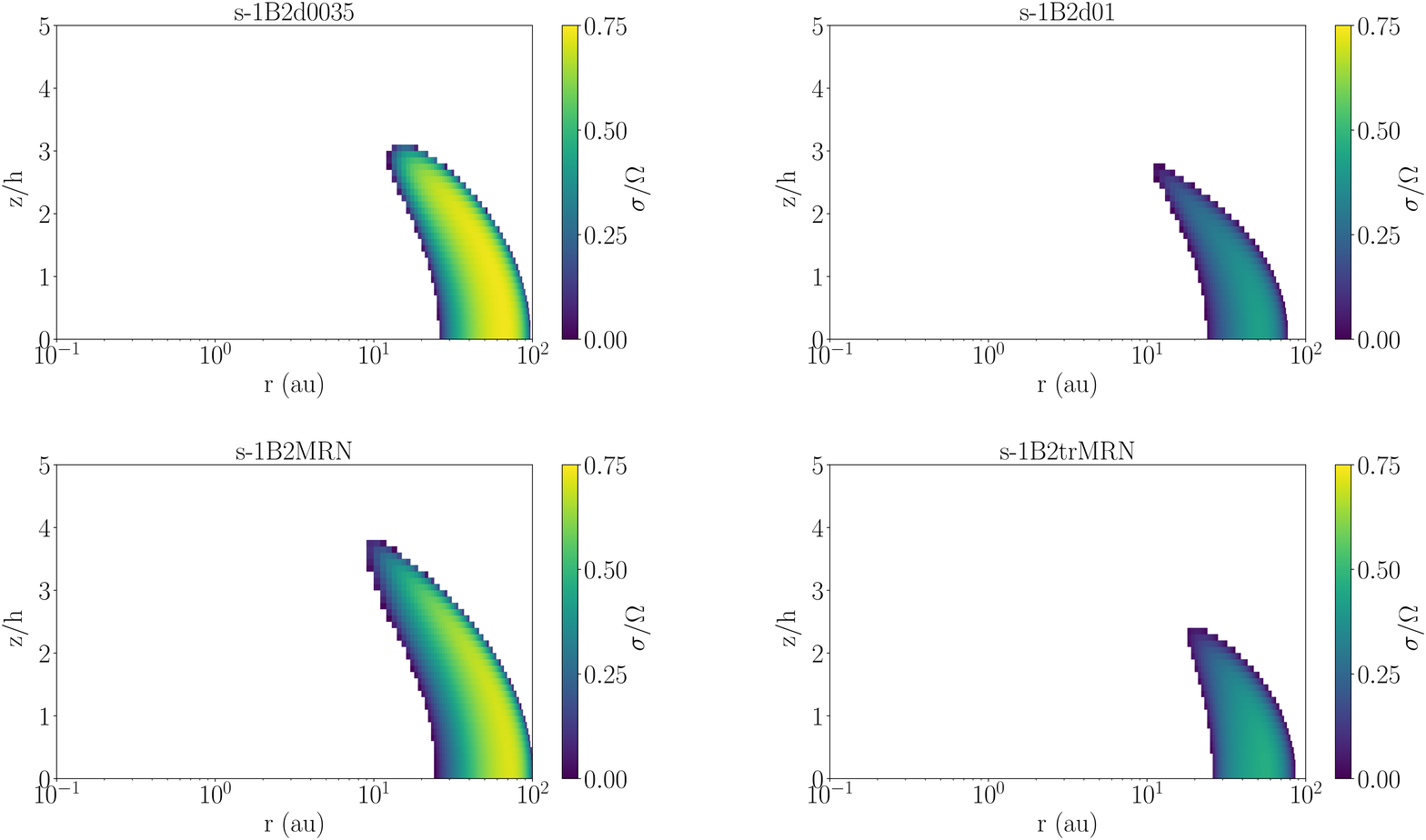}
    \caption{           
            As Fig.~\ref{fig:B1-s1} but for different models s-1B2d0035 (top left), s-1B2d01 (top right), s-1B2MRN  (bottom left), s-1B2trMRN  (bottom right). 
            The models have the same parameters of  $B=0.01$\,G and $s=-1$, but are different dust models. 
            }
    \label{fig:B2-s-1}
\end{figure}

The MRI growth rate for models with $B=0.01$\,G and $s=-1$ is plotted in Figure~\ref{fig:B2-s-1}. 
The MRI active area for the models with $B=0.01$\,G is larger than for the model with $B=0.1$\,G. 
As described above, the ambipolar diffusion and Hall coefficients decrease as the magnetic field strength decreases, and thus the MRI tends to grow with a weak magnetic field. 
In the models in Figure~\ref{fig:B2-s-1}, the growth rate $\sigma$ is larger in the model with relatively small-sized dust grains than in the model with relatively large-sized dust grains, while the growth area of MRI is almost the same among the models. 

In this subsection, we show the MRI growth rate for four cases (Figs.~\ref{fig:B1-s1}, \ref{fig:B2-s1}, \ref{fig:B1-s-1} and \ref{fig:B2-s-1}) with different strengths and directions of the magnetic field. 
Among the cases, the MRI active area is largest when the magnetic field is as weak as $B=0.01$\,G and the magnetic field  is anti-parallel to the rotation vector. 
However, even in this case, the MRI active region is limited to $r>20$--$30$\,au, and the growth rate strongly depends on the dust size distribution. 

\section{Discussion}
\subsection{Configuration and Strength of Magnetic Field in Disk}
As described above,  the magnetic field strength is a factor in whether MRI develops in a weakly ionized disk 
because the ambipolar diffusion and Hall coefficients depend on the magnetic field strength. 
It should be noted that MRI does  not develop  with a strong magnetic field even when the ideal MHD approximation is applicable 
\citep{1998RvMP...70....1B,2012MNRAS.423.1318O,2012A&A...548A..76M,2014prpl.conf..411T}.  
In this study,  we adopted a uniform magnetic field, as justified in \citet{2016ApJ...830L...8H}, while a non-uniform magnetic field distribution may give a different outcome. 
In addition, for simplicity, we only assumed the vertical component of the magnetic field in the disk. 
However, we may need to consider other components of the magnetic field such as a toroidal component to precisely investigate the growth rate of MRI.
It should be noted that \citet{2011ApJ...742...65O} showed that the growth rate of the MRI is determined mainly by the strength of the vertical component of  the magnetic field. 
In addition, we only considered the circumstellar disk in the main accretion phase during which the vertical component of the magnetic field dominates 
the other components, because ohmic dissipation and ambipolar diffusion stretch the magnetic field lines in the vertical direction 
\citep{2015ApJ...801..117T}.

During the main accretion phase, magnetic flux is introduced by mass accretion with flux freezing, while it dissipates by non-ideal MHD effects. 
Thus, when accretion dominates dissipation, the magnetic flux moves inward and vice versa. 
In addition,  it should be stressed that the disk is not an isolated system in  the main accretion phase, during 
which the magnetic field lines are connected from the disk to the infalling envelope (or remnant of the star-forming cloud core). 
Thus, it is very difficult to estimate the distribution and strength of  the magnetic field. 

In addition, there is a caveat about the magenetic field strength at large radii.
In this study, adopting a constant magnetic field  strength of 0.1 and 0.01\,G, we estimated the diffusion coefficients to investigate whether MRI grows in the range of 0.1-100\,au.
However, in reality, the distribution of the magnetic field is determined by the physical processes described above. 
Meanwhile, both observation and simulation imply that the size of the circumstellar disks around Class 0 protostars is limited to $\lesssim 10-20$\,au. 
Our analysis can be applicable within $\lesssim 10-20$\,au where the assumption of a constant magnetic field would be  appropriate \citep{2016ApJ...830L...8H}.
On the other hand, our estimate of the growth rate of MRI may not be applicable at large radii of $\gg20$\,au because magnetic field strength may not be as strong as 0.1 and 0.01\,G there. 
We plan to study the time evolution of the magnetic field distribution in  the main accretion phase using one- or two-dimensional simulations to better 
understand MRI in the very early star formation phase. 

\subsection{Dust Growth and Sedimentation}
As shown in \S\ref{sec:results}, MRI does not develop  in the inner disk region in any model. 
If no other instabilities (including MRI) occur in the disk, it is expected that the dust grains will settle 
into the equatorial plane without stirring (or without turbulence). 
Then, dust growth should be promoted on the equatorial plane. 
In this study, we assumed a dust-to-gas ratio of $f_{\rm dg} =  0.01$ in the whole disk region.
However, the dust-to-gas ratio possibly decreases as dust settlement or dust growth proceeds, except for at the equatorial plane. 
The abundance of the dust grains greatly affects the abundance of ions and electrons.
The ionization degree and diffusion coefficients also depend on the amount of dust. 
Therefore, the growth rate of MRI should change according to the amount of dust. 

\begin{figure}
    \centering
    \includegraphics[clip, width=160mm]{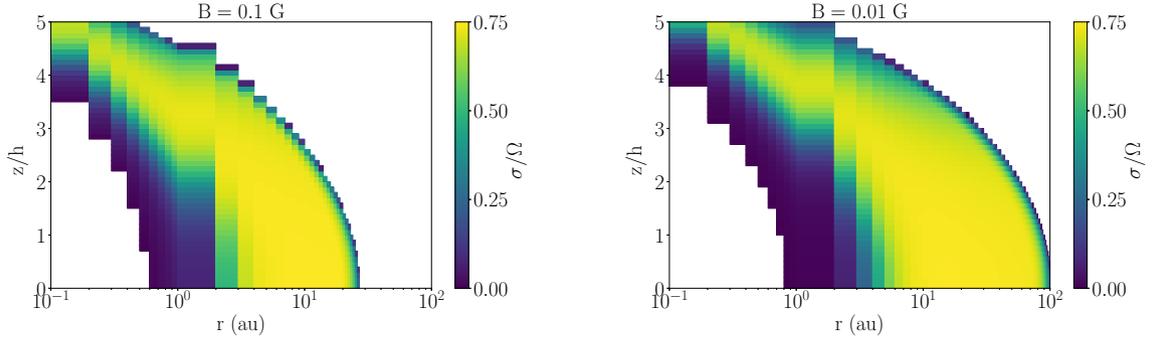}
    \caption{
        Growth rate of MRI on the $r$--$z$ plane for models with $f_{\rm dg}=0$. 
        A uniform magnetic field of $B=0.1$\,G (left) and 0.01\, (right) is adopted.  
        The $z$-direction is normalized by the scale height at each radius $r$. 
        The color denotes the growth rate of the fastest growth mode normalized by the Keplerian angular velocity. 
        }
    \label{fig:Iso-nodust2}
\end{figure}

As an extreme case, we calculated the growth rate of MRI with $f_{\rm dg} = 0$.
The results are shown in Figure~\ref{fig:Iso-nodust2}, in which uniform magnetic fields of $B=0.1$\,G (left) and 0.01\,G (right) are adopted with $s=1$.
In the figure, although MRI is still suppressed by ohmic dissipation within $\lesssim 1$\,au,
the MRI active area is widely distributed in the range $5\,{\rm au} \lesssim r \lesssim 100\,{\rm au}$ even  on the equatorial plane. 
In reality, the dust settlement and MRI should occur at the same time.
Realistic non-linear simulations of the early star formation stage are necessary to more precisely understand the growth of MRI. 

\subsection{Other Mechanisms of Angular Momentum Transfer}
When the disk is stable against MRI, other mechanisms of angular momentum transfer are necessary to realize accretion from the disk onto the protostar. 
In the main accretion phase, mass is continuously supplied from the infalling envelope.
Thus, without an efficient mechanism of angular momentum transfer, gravitational instability in the disk  should naturally occur, promoting mass accretion onto the protostar  
\citep{2017ApJ...835L..11T}. 
In addition, even if MRI does not occur,  the magnetic field plays an important role in transporting the angular momentum in  the disk. 
The infalling envelope can apply braking to the rotationally supported disk through the magnetic field lines (i.e. magnetic braking).
In addition, the angular momentum is conveyed by magnetically driven winds
\citep{2009ApJ...691L..49S,2017ApJ...845...75B,2017A&A...600A..75B,2018ApJ...865...10S}. 
Furthermore, the Hall shear instability should play a role in transporting the angular momentum in the disk
\citep{2014A&A...566A..56L,2015MNRAS.454.1117S,2017ApJ...836...46B}. 
The vertical shear instability may be also important for the angular momentum transport
\citep{2013MNRAS.435.2610N,2014A&A...572A..77S,2015ApJ...811...17L}. 

Each mechanism of angular momentum transfer has been investigated in different studies with different settings. 
Magnetic braking and large-scale outflows were investigated in core collapse simulations covering the whole disk. 
Disk wind and HSI (and MRI) were investigated in global disk simulations that cover a large part of the protoplanetary disk but do not cover or ignore the infalling envelope
\citep{2017ApJ...845...75B,2017A&A...600A..75B,2020ApJ...896..126G}. 
MRI and vertical shear instability were investigated in  local disk simulations that covered only a very small part of the disk
\citep{1995ApJ...440..742H,2013ApJ...767...30B,2013MNRAS.435.2610N}. 
The spatial resolution and timescale covered in the simulations differ considerably. 
The settings of the linear analysis were also different, depending on which phenomenon (MRI, HSI, VSI) was being focused on. 
In this study, although we investigated the possibility of the growth of MRI in an early star formation stage, we need to further investigate other effects of angular momentum transfer in future studies.

\subsection{Caveats of Our Disk Model}
\label{sec:caveat}
As described in \S\ref{sec:diskmodel}, the disk model used in this study is analytically constructed with  Toomre Q parameter, in which the density and temperature distributions are fixed.   
With  the disk model, we investigated whether MRI develops using a linear analysis and showed that the MRI active region is severely limited in the disk.

Meanwhile, this work does not actually calculate or simulate the growth of MRI in a gravitationally unstable disk.
In other words, the disk evolution is not determined by a two or three dimensional simulation.
Thus, we would ignore some factors seen in the simulations such as  temporal change in the disk profile and  perturbation attributed to  gravitational instability.
Thus, the actual behavior  of MRI in a gravitationally unstable disk could be different from our results. 
 
For example, \citet{2018MNRAS.474.2212R,2019MNRAS.482.3989R} investigated  MRI in gravitationally unstable disks with their MHD simulations. 
They calculated MRI in such disks embedded in a shearing box.
They  showed that perturbations caused by gravitational instability interact with MRI and can suppress the growth of  MRI.
However, they only included the ohmic dissipation term and ignore both the ambipolar diffusion and Hall terms.
In addition, they assumed zero net flux, while non-zero net flux is adopted in this study (\S\ref{sec:crit_mri} and \S\ref{sec:magneticfield}). 
Thus, the growth of MRI is suppressed in both studies, while we can not fairly compare our results with \citet{2018MNRAS.474.2212R,2019MNRAS.482.3989R}. 
We need to investigate the growth of MRI with more realistic settings with numerical simulations in future studies.

\section{Summary}
We evaluated whether MRI can grow in a circumstellar disk during the main accretion phase.
First, we modeled a disk regulated by disk gravitational instability according to the prescription adopted in our previous study. 
When constructing the disk model, we assumed a uniform magnetic field with different strengths ($B$ = 0.1 and 0.01\,G) and different directions (magnetic field parallel to or anti-parallel to the angular momentum vector of the disk).
Then, we calculated the chemical reactions and derived the electric conductivities and diffusion coefficients of the magnetic field in each area  of the disk, in which we adopted four different dust models (two different single sized dust models, and MRN and tr-MRN size distribution dust models). 
Finally, we estimated the growth rate of MRI using the diffusion coefficients of the magnetic field, and determined whether MRI grows using a linear analysis of the MRI including the non-ideal MHD terms of ohmic dissipation, ambipolar diffusion and the Hall effect. 

Our calculations showed that MRI is suppressed both in the inner and outer disk regions with a relatively strong magnetic field ($B=0.1$\,G) when the magnetic field is parallel to the disk angular momentum. 
The ohmic dissipation stabilizes the disk against MRI in the inner disk region, while ambipolar diffusion suppresses the MRI in the outer disk region. 
In the intermediate region of the disk, Hall-shear instability can promote MRI growth. 
On the other hand, when the magnetic field  is anti-parallel to the disk angular momentum, the growth of MRI is suppressed over the whole  region of the disk. 

When the size of the dust grains is small, the MRI growth area is widest when the magnetic field is as weak as $B=0.01$\,G 
and the magnetic field  is anti-parallel to the disk angular momentum. 
However, even in this case, MRI grows only in a limited range of $\sim 30$--$50$\,au. 
Although the MRI active region appears in the limited range of $\gtrsim 30$\,au on the equatorial plane, there is no MRI active region within $\lesssim 30$\,au. 
Both the observations and simulations indicate that the size of the Keplerian disk around Class 0 protostars is limited to $\lesssim 10$--$20$\,au with a few exceptions. 
Some simulations showed that the Keplerian disk exponentially grows as the infalling envelope dissipates after the  magnetic braking is alleviated \citep[e.g.][]{2011PASJ...63..555M}. 
Thus, a disk size of $\lesssim 10$--$20$\,au is considered to be realistic for a disk during the main accretion phase. 
Our study showed that MRI cannot grow within $\lesssim 30$\,au except for the disk inner edge.
Thus, we conclude that MRI does not significantly influence  the evolution of the disk during the main accretion phase. 

\section*{Acknowledgements}
We thank the referee for very useful comments and suggestions on this paper. 
The present study was supported  by JSPS KAKENHI Grants (JP17H02869, JP17H06360, JP17K05387, JP17KK0096: MNM).

\section*{Data Availability}
The data underlying this article are available in the article.

\bibliographystyle{mnras}
\bibliography{example} 

\clearpage

\clearpage

\clearpage

\clearpage

\clearpage

\clearpage

\clearpage

\clearpage

\clearpage

\clearpage

\bsp	
\label{lastpage}
\end{document}